\documentclass[aoas,preprint]{imsart}
\pdfoutput=1
\setattribute{journal}{name}{}

\RequirePackage[OT1]{fontenc}
\RequirePackage{amsthm,amsmath}
\RequirePackage{natbib}%[authoryear][numbers][imsart-nameyear]
\RequirePackage{xr-hyper} 
\RequirePackage[colorlinks,citecolor=blue,urlcolor=blue]{hyperref}

\RequirePackage{stmaryrd}
\RequirePackage{graphicx}
\RequirePackage{tabularx}
\RequirePackage{lipsum}
\RequirePackage[thinlines]{easytable}
\RequirePackage{amssymb}
\RequirePackage{bbm}
\RequirePackage{array}
\RequirePackage{verbatim}
\RequirePackage{enumerate}
\RequirePackage{amsmath, amsfonts, amsthm, amscd}       %formattazione matematica
\RequirePackage{fancybox}
\RequirePackage[T1]{fontenc}
\newlength{\querylen}
\newcommand{\R}{\mathbb{R}}

\newcommand{\I}{\mathbb{I}}
\newcommand{\1}{\mathbbm{1}}
\newcommand{\Pn}{\mathcal{P}_n}

\theoremstyle{plain}

\newtheorem{rmk}{Remark}

\newtheorem{?}{Question}

\startlocaldefs
\numberwithin{equation}{section}
\theoremstyle{plain}

\endlocaldefs
\begin{document}

\begin{frontmatter}
\title{Bayesian Complementary Clustering, MCMC and Anglo-Saxon placenames}%\thanksref{T1}} for Data Association
\runtitle{Complementary Clustering for Anglo-Saxon placenames}
%\thankstext{T1}{Footnote to the title with the ``thankstext'' command.}

\begin{aug}
\author{\fnms{Giacomo} \snm{Zanella}\thanksref{t1}\ead[label=e1]{g.zanella@warwick.ac.uk}}%t1,,m1

%\thankstext{t1}{Some comment}
\thankstext{t1}{Supported by EPSRC through a PhD position under the CRiSM grant EP/D002060/1}
\runauthor{G. Zanella}

\affiliation{University of Warwick}%\thanksmark{m1}

\address{Department of Statistics\\
University of Warwick\\
Coventry, CV4 7AL\\
United Kingdom\\
\printead{e1}}
\end{aug}

\begin{abstract}
%We consider a dataset including locations and placenames of 1316 Anglo-Saxon settlements dated between 750 and 850 AD. Historians hypothesize that such settlements are organized in administrative clusters involving complementary placenames. We investigate the evidence for such an hypothesis by developing a Bayesian Random Partition Model based on clusters formed by points of different types (complementary clustering). 

%As a result we obtain an intractable posterior distribution on the space of matchings contained in a k-partite hypergraph. We apply the Metro$\-$po$\-$lis-Hastings (MH) algorithm to sample from this posterior. We consider the problem of choosing an efficient MH proposal distribution and we obtain consistent mixing improvements compared to the choices found in the literature. Simulated Tempering techniques can be used to overcome multimodality and a multiple proposal scheme is developed to allow for parallel programming. Finally, we discuss results arising from the careful use of convergence diagnostic techniques.

%This allows us to study the Anglo-Saxon placenames locations dataset. Without strong prior knowledge, the model allows for explicit estimation of the number of clusters, the average intra-cluster dispersion and the level of interaction among placenames. The results suggest modest support for the hypothesis of organization of settlements into administrative clusters based on complementary names.

%\vspace{5cm}

Common cluster models for multi-type point processes model the aggregation of points of the same type. In complete contrast, in the study of Anglo-Saxon settlements it is hypothesized that administrative clusters involving complementary names tend to appear. We investigate the evidence for such an hypothesis by developing a Bayesian Random Partition Model based on clusters formed by points of different types (complementary clustering).

As a result we obtain an intractable posterior distribution on the space of matchings contained in a k-partite hypergraph. We apply the Metro$\-$po$\-$lis-Hastings (MH) algorithm to sample from this posterior. We consider the problem of choosing an efficient MH proposal distribution and we obtain consistent mixing improvements compared to the choices found in the literature. Simulated Tempering techniques can be used to overcome multimodality and a multiple proposal scheme is developed to allow for parallel programming. Finally, we discuss results arising from the careful use of convergence diagnostic techniques.

This allows us to study a dataset including locations and placenames of 1316 Anglo-Saxon settlements dated approximately around 750-850 AD. Without strong prior knowledge, the model allows for explicit estimation of the number of clusters, the average intra-cluster dispersion and the level of interaction among placenames. The results support the hypothesis of organization of settlements into administrative clusters based on complementary names.

\end{abstract}


\end{frontmatter}

\section{Introduction}
\subsection{The historical problem}
The starting point of this work is a dataset supplied by Professor John Blair of Queen's College, Oxford. The dataset consists of the locations and placenames of 1316 Anglo-Saxon settlements dated approximately around 750-850 AD (dataset fully available in \ref{supp:data_codes}). %, for illustrations see \ref{supp:tables_plots} or Figure \ref{fig:grayplot}).
In the dataset there are 20 different kinds of placenames in total. Placenames form an important source of information regarding the Anglo-Saxon civilization and are intensively studied by the historical community (see for example \citet{Gelling2000} and \citealp{Jones2012}).

In particular, the placenames included in this dataset are often described as \emph{functional} placenames, as they were probably used to indicate specific functions or features of their corresponding settlements. %historians think that the placenames included in the dataset under consideration were used to indicate specific functions or features of each settlement, and thus are often referred to as \emph{functional} placenames.
For example $Burton$ is thought to label fortified settlements having a military role, $Charlton$ the settlements of the peasants and $Drayton$ the settlements dedicated to portage.
%Nonetheless we note that there is a lot of uncertainty and controversy regarding the meaning of placenames (even the apparently obvious ones).

\begin{figure}[h!]
\centering  \includegraphics[width=\linewidth]{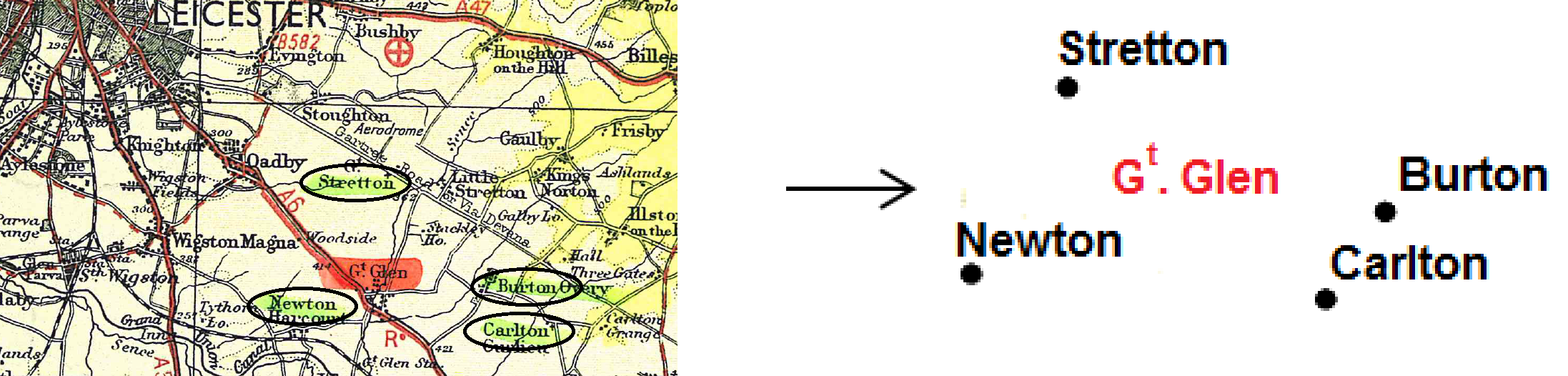}
\caption{A cluster of four Anglo-Saxon settlements (highlighted in green and circled) in the region of Great Glen (highlighted in red).}\label{fig:blair1}
\end{figure}

Moreover historians expect the settlements in this dataset (especially those having one of the placenames underlined in Table \ref{Summary_Table}) to have been formed approximately at the same time and in the same context (specifically, royal administration in the period c.750-850).
This suggests that there could be some coherence in the distribution of such placenames.
In particular Professor Blair's hypothesis is that those settlements were not independent units but rather that they were organized into administrative clusters (or districts) where placenames were used to indicate the role of each settlement within the district.
According to this hypothesis such clusters would tend to involve a variety of complementary
placenames in each of them. For example Figure \ref{fig:blair1} indicates a plausible administrative cluster made of four settlements, with, for example, a settlement dedicated to military functions (\emph{Burton}) and one dedicated to agriculture (\emph{Carlton}).
%Note that historians expect those settlements to have been formed approximately at the same time and in the same context (specifically, royal administration in the period c.750-850).
%If the distribution of such placenames was showing some coherence, this would be in accordance with such an hypothesis.
%A coherence in the distribution of such placenames would be connected to the fact that those settlements may have been formed approximately at the same time and in the same context (specifically, royal administration in the period c.750-850).% So our initial expectation is that that there is likely to be some coherence in the distribution of these names.

The objective of our statistical approach to the study of settlements names and geographical locations is to address the following questions: is there statistical support for Blair's hypothesis? What is the typical distance between settlements in the same cluster? How many settlements are clustered together and how many are singletons? Which placenames tend to cluster together? Can we provide a list of those clusters which are more strongly supported by the analysis?

Our intention is to provide a useful contribution to historical research on this topic based on a quantitative approach, bearing in mind the scarcity of textual evidences regarding the Anglo-Saxon period. 
%Given the scarcity of textual evidences regarding the Anglo-Saxon period, 
Since there is a lot of uncertainty and controversy regarding the meaning of placenames, even the apparently obvious ones, we should try to be fairly neutral from the historical point of view, avoiding strong assumptions on the functions of placenames and relationships among them. This will help our statistical analysis to be a genuine contribution to the ongoing historical debate on this topic.

We note that there has already been statistical work related to Anglo-Saxon placenames. In particular Keith Briggs did various works on this topic (see \url{http://keithbriggs.info/place-names.html} for a full list). Nevertheless both the historical questions considered and the statistical meth$\-$odologies used are substantially different from ours.

\subsection{Modeling approach}By considering the placenames as marks attached to points, we model our data as the realization of a $k$-type point process (also called $k$-variate point process), where $k$ is the number of different placenames available (see \citealp{Baddeley2010}). We can view our problem as a clustering problem based on aggregations of points of different types. In fact we seek a \textit{complementary clustering}: each cluster may contain at most one settlement for each placename. This simplifying requirement is motivated by the assumption that each placename represents a different administrative function (role) within the cluster.%requirement is imposed as a result of consultation with the historians involved in the project and This requirement is imposed as a result of consultation with the historians involved in the project and is motivated by the assumption that each placename represents a different administrative function (role) within the cluster.

Our intention is to perform explicit inferences on the partition of settlements into clusters. As with hierarchical models, it would be desirable to analyze the dataset all at once, so at not to loose statistical power, and also to provide inferences at the single cluster level to facilitate visualization and historical interpretation of the results of the analysis. 

We employ Random Partition Models (RPMs), often used in the Bayesian Nonparametric literature (e.g.\ \citealp{Lau2007}), as they permit natural inferences on the cluster partition and they have enough flexibility to allow specification of a useful model for complementary clustering.

Standard approaches for point process cluster modeling, like the Log-Gaussian Cox Processes (see \citealp[ch.3]{Lawson2010}) or the Neyman-Scott model (e.g.\ \citealp{Loizeaux2001}), are not appropriate here, as such models usually provide inferences on the cluster centers or on the point process intensity, while we seek explicit inferences on the cluster partition. Moreover standard cluster methods for marked point process consider the marks as an additional dimension and search for aggregations of points with similar marks. In complete contrast, we seek for aggregations of points of different types.

\citet{Diggle2006} seek evidence for repulsion among points of different types in a bivariate spatial distribution of amacrine cells. They use a pairwise interaction model, which has theoretical limitations which prevent its use for clustering. While this approach could be extended to our case by using area-interaction point processes, which can model clustering \citep{Baddeley1995}, it would not provide us with explicit estimates of the cluster partition and it would not easily allow complementary clustering specification (at most one point of each type in each cluster).

Multi-target tracking involves the Data Association problem, that is to group together measurements recorded at different time intervals to create objects tracks (e.g \citealp{Oh2009}). This problem is similar to the problem of performing complementary clustering of a $k$-type point process. In Data Association problems, however, the interest is to find the best association, while we are interested in assessing the strength of clustering and the level of interaction between different placenames, and in quantifying the uncertainty of our estimates. In fact the modeling aspects we have to be careful about are different from the ones of Data Association problems, while the computational challenges are similar (see Sections \ref{sec:intractability} and \ref{sec:MCMC}).

\subsection{Organization of the paper}
In Section \ref{sec:prel_an_and_k_fun} we perform preliminary analysis of the dataset, testing whether there is a significant clustering interaction between points of different types by using common Spatial Statistics tools such as K-cross functions.
In Section \ref{sec:AS_model} we define a RPM for complementary clustering and discuss appropriate prior distributions for the cluster partition (see also Section \ref{sensitivity-sec:dir_mult_model} of \ref{supp:sensitivity}). The resulting model leads to an intractable posterior distribution.
We express such a posterior in terms of matchings contained in hypergraphs. We thus link the problems of sampling from the posterior and finding the posterior mode to the more classical problems of Data Association and Optimal Assignment.
In Section \ref{sec:MCMC} we design a Metropolis-Hastings algorithm to obtain approximate samples from the posterior. We carefully consider the problem of choosing an efficient proposal distribution, we explore the use of Simulated Tempering to overcome multimodality and we develop a multiple proposal scheme to allow for parallel computation.
In Section \ref{sec:application} we analyze the Anglo-Saxon placename location data with our RPM, using the algorithm of Section \ref{sec:MCMC}. The results support the hypothesis of settlements being organized into administrative clusters and give explicit inferences of various quantities of historical interest.
Finally in Section \ref{sec:discussion} we discuss future directions of research. Supplementary material includes extensive calculations, additional tables and plots, the settlements dataset and R codes to perform the data analysis. 

\section{Preliminary analysis of the Anglo-Saxon settlements dataset}\label{sec:prel_an_and_k_fun}
We describe the Anglo-Saxon settlements dataset supplied by Prof.\ John Blair and the data cleaning operations that we carried out. We then perform preliminary analysis on the resulting point pattern using Spatial Statistic tools.

\subsection{Format of the dataset}\label{subsec:Prel_an}
The dataset (available in \ref{supp:data_codes}) is made of 20 different groups, each of which contains the list of settlements having one of the 20 placenames (see Table \ref{Summary_Table}).
The historians involved in the project expect the clustering behaviour to involve in particular 13 of those placenames, indicated in Table \ref{Summary_Table}. We refer to the settlements relative to those 13 placenames as \emph{reduced dataset}, and to all the settlements recorded as \emph{full dataset}. We will perform the analysis on both datasets.

For each settlement the following variables are given: County, place, Parish or Township, grid ref, date of first evidence (see Table \ref{table:example}).
\begin{table}[h!]
  \centering\begin{tabular}{|c|c|c|c|c|}
              \hline
              &&PARISH OR&GRID&DATE OF\\
              COUNTY&PLACE& TOWNSHIP&REF&FIRST\\
 		  & & & &EVIDENCE \\\hline
              BRK&Bourton&Bourton&SU 230870&c. 1200\\
              BUC &Bierton&Bierton with Broughton&SP 836152&DB\\
              BUC &Bourton&Buckingham&SP 710333&DB\\
              CHE&Burton&Burton (T)&SJ 509639&DB\\
              CHE &Burton&Burton (T)&SJ 317743&1152\\
              CHE &Buerton&Buerton (T)&SJ 682433&DB\\
              \hline
            \end{tabular}
  \caption{Data available regarding the first $6$ settlement with the name $Burton$. The acronym \textrm{DB} stands for Domesday Book, compiled in 1086.}\label{table:example}
\end{table}

The locations are expressed through the Ordnance Survey (OS) National Grid reference system. A set of OS National Grid coordinates, like $SU 230870$, identify a $100\hbox{m}\times 100\hbox{m}$ square on a grid covering Great Britain. Some locations have just 2 letters and 4 digits (e.g.\ $SU 2387$) and they identify a $1\hbox{km}\times 1\hbox{km}$ square, and some have a letter $c$ in front of them (e.g. $c. SU 2387$) to indicate that the location is less accurate (see Table \ref{Summary_Table} for amounts).

\begin{table}[h!]
  \centering
\resizebox{0.96\columnwidth}{!}{
\begin{tabular}{|l|c|c|c|c|}
              \hline
              % after \\: \hline or \cline{col1-col2} \cline{col3-col4} ...
              Placenames & total  & $\#$ of settlements & $\#$ of couples & $\#$ of couples\\
               & number & with less precise & (as classified   & (as classified  \\
               &  & location & by historians)  & by proximity)  \\\hline
              \underline{\textbf{Aston/Easton}} & 90 & 0 & 1 & 8\\
              Bolton & 17 & 1 & 1 & 0 \\
              Burh-Stall & 29 & 2 & 1 & 0\\
              \underline{\textbf{Burton}} & 108 & 2 & 1 & 7\\
              Centres & 46 & 0 & 0 & 0\\
              \underline{\textbf{Charlton/Charlcot}} & 98 & 3 & 7 & 1\\
              Chesterton & 9 & 0 & 0 & 0\\
              Claeg & 84 & 13 & 0 & 5\\
              \underline{\textbf{Draycot/Drayton}} & 55 & 1 & 0 & 2\\
              \underline{\textbf{Eaton}} & 33 & 1 & 1 & 5\\
              \underline{\textbf{Kingston}} & 71 & 1 & 1 & 1\\
              \underline{\textbf{Knighton}} & 26 & 1 & 0 & 0\\
              Newbold & 34 & 3 & 1 & 0\\
              \underline{\textbf{Newton}} & 191 & 5 & 4 & 5\\
              \underline{\textbf{Norton}} & 74 & 1 & 8 & 1\\
              \underline{\textbf{Stratton}} & 37 & 0 & 5 & 0\\
              \underline{\textbf{Sutton}} & 101 & 2 & 4 &5\\
              Tot & 77 & 17 & 1 & 1\\
              \underline{\textbf{Walton/Walcot}} & 51  & 4 & 1 & 0\\
               \underline{\textbf{Weston}} & 85 & 3 & 3 & 2\\\hline
              Total & 1316 & 60 & 40 & 43\\
              \hline
            \end{tabular}
}
\caption{ Number of settlements in the Anglo-Saxon placenames location dataset supplied by Prof.\ Blair. The historians expect the clustering behaviour mainly to involve 13 of those placenames (underlined and emboldened in this table). Settlements with less precise locations (third colums) are settlements whose location is given with 1 km accuracy, rather than 100 m, or having a more uncertain location (see Section \ref{subsec:Prel_an}). The term ``couples'' (last two columns) refers to multiple records of the same settlements (see Section \ref{sec:data_cleaning} for discussion). The ``total number'' column refers to the count after merging the couples classifieds by historians.}
\label{Summary_Table}
\end{table}

\subsection{Data cleaning and data assumptions}\label{sec:data_cleaning}
%Our analysis is concerned with the placenames (variable ``place'') and the geographical locations (variable ``Grid reference''). In order to use our model we need to code the placenames as $k$ types and convert the locations to two-dimensional Euclidean coordinates. Such 

Our analysis is concerned with placenames (variable ``place'') and geographical locations (variable ``Grid reference'').
We convert the data to a k-type point process form as described below.
Such data cleaning process entails historical assumptions on the dataset and thus we have been guided by the judgment of the subject-specific historians involved in this project in doing so.

\textit{Placenames:} we express the variable ``place'' as a categorical variable with $k$ possible values (i.e.\ $k$ types).
By doing so we ignore minor variation in placenames. For example we consider the settlements of Table \ref{table:example} as having placename $Burton$: their actual recorded placenames vary amongst \textit{Burton, Bourton, Bierton, Buerton}.

Four groups (out of 20) are made up of two subgroups each with similar placenames: $Aston/Easton$, $Charlton/Charlcot$, $Drayton/Draycot$ and $Walton/Walcot$. We consider such subgroups to be the same, for example $Charlton$ and $Charlcot$ are treated as the same placename.

\textit{Locations:} we convert OS National Grid coordinates to two-dimensional Euclidean coordinates and each settlement is assumed to be located at the center of the corresponding OS National Grid square.

\textit{``Multiple'' records:} it is sometimes indicated in the original dataset that some couples (or triples) of settlements, with same placename and very close locations, have to be considered as multiple records of the same settlement.
We replaced such couples (or triples) of settlements with one settlement located at their midpoint.
Moreover there are some other pairs of records having very close locations and the same placename (see Table \ref{Summary_Table} for amounts).
It is primarily a matter of historical interpretation whether these couples have to be considered as single settlements. We performed the analysis under both hypothesis (keeping them separated and merging them) without seeing significant change in the results. The analysis presented here is made with those settlements merged together (3 km is the threshold distance below which we identify two records of settlements with the same placename).

\textit{Observation region $W$:}
a point processes realization consists of points locations \emph{and} of the region $W$ where the points have been observed.
Indeed both the K-cross function analysis of Section \ref{subsec:Kcross} and the Bayesian analysis of Section \ref{sec:AS_model} will use informations about $W$.
In our case we define $W$ as Great Britain (coastline obtained from the $mapdata$ $R$ package \citealp{mapdata}) cropping the region where the point process intensity $g$ falls below a certain threshold, approximately at the borders between England-Scotland and England-Wales. We also added a small buffer zone of $3$ km around the region to include the few points that were falling outside the region (e.g. because the coastline has moved or because the location was inaccurate). See Figure \ref{fig:grayplot} for a plot of the region.
%The intensity $g$ used to perform such an operation is estimated using standard Gaussian kernel smoothing (bandwidth chosen according to the cross-validation method, see for example \citet{Diggle2003}, p.115-118). Finally, we added a small buffer zone of $3$ km around the region to include the few points that were falling outside the region (e.g. because the coastline has moved or because the location was inaccurate). In \ref{supp:sensitivity} we test how sensitive our analysis

\subsection{K-cross function analysis}\label{subsec:Kcross}
Second moment functions are a useful tool to investigate interpoint interaction (e.g. \citealp{Stoyan1987}). In particular, given a multi-type point pattern, bivariate (or cross-type) K-functions provide good summary functions of the interaction across points of different types. The bivariate K-function $K_{ij}(r)$ is the expected number of points of type $j$ closer than $r$ to a typical point of type $i$, divided by the intensity $\lambda_j$ of the type $j$ subpattern of points $\textbf{x}_j$ (e.g. \citealp[Sec. 6]{Baddeley2010}). For testing and displaying purposes we define a single summary function, a multi-type K-function $K_{cross}(r)$, as the weighted average of $K_{ij}(r)$ for $i\neq j$, where the weights are the product of the intensities $\lambda_i\lambda_j$.

Classical $K$-functions, however, rely strongly on the assumption that the point pattern is stationary, which is not appropriate for our dataset. Therefore we use the inhomogeneous version of the K-functions, where the contribution coming from each couple of points is reweighted to take into account for spatial inhomogeneity \citep{Baddeley2000}. Standard estimates of the inhomogeneous bivariate $K$-functions $\hat{K}_{ij}$ are obtained using the $spatstat$ $R$ package \citep{spatstat}.

\subsubsection{Null hypothesis testing}\label{sec:null_hypothesis}
In order to test whether the interaction shown by $K$-functions is significant or not we need to define a null hypothesis (representing no-interaction among placenames). Section 8 of \citet{Baddeley2010} describes three classical null hypotheses for multivariate point processes: random labeling (given the locations the point types are i.i.d.), Complete Spatial Randomness and Independence (CSRI, the locations arise from a uniform Poisson point process and the point types are i.i.d.) and independence of components (points of different types are independent). The random labeling and the CSRI hypotheses are unrealistic assumptions for our dataset because our point pattern is clearly not stationary and the distribution of placenames is not spatially homogeneous (some placenames are more concentrated in the South, some in the North and so on). The independence of components hypothesis is realistic but, in order to test it, stationarity of the points pattern is usually assumed. Instead we define the following no-interaction null hypothesis: each subpattern of points $\textbf{x}_j$ is an inhomogeneous Poisson point process (with intensity function $\lambda_j(\cdot)$ potentially varying over $j$).
Note that a more realistic null hypothesis would include repulsion among points of the same type. % because similar placenames are unlikely to be used in close proximity.
In Section \ref{sensitivity-sec:strauss} of \ref{supp:sensitivity} we implement such a null hypothesis using Strauss point-processes. The results are very similar to the ones presented here and require additional tuning of various parameters.%in Figure \ref{fig:k_cross_env}.

Given the null hypothesis we perform the following approximate Monte Carlo test.
First we estimate the intensities $\lambda_j(\cdot)$ with $\hat{\lambda}_j(\cdot)$ (see Figures \ref{plots-fig:density_estimate} and \ref{plots-fig:density_estimates} of \ref{supp:tables_plots}) obtained through standard Gaussian kernel smoothing with bandwidth chosen according to the cross-validation method (e.g. \citealp[p.115-118]{Diggle2003}), and edge correction performed according to \citet{Diggle1985}.
Secondly we sample 99 independent multivariate inhomogeneous Poisson point patterns according to $\big\{\hat{\lambda}_j(\cdot)\big\}_{j=1}^k$.
Finally we use those samples to plot simulation envelopes and to perform a deviation test with significance $\alpha=0.05$ using as a summary function a centered version of the $L$-function $\hat{L}_{cross}(r)=\sqrt{\frac{\hat{K}_{cross}(r)}{\pi}}$ for $r\in(0,r_{max})$, with $r_{max}=15$km. 
The deviation test \citep{Grabarnik2011} summarizes the summary function with a single value $D=\max_{r\in(0,r_{max})}\hat{L}_{cross}(r)-\mathbb{E}[\hat{L}_{cross}(r)]$ and compares it to the ones obtained from the $99$ simulated samples.
\begin{figure}[h!]
\includegraphics[width=\textwidth]{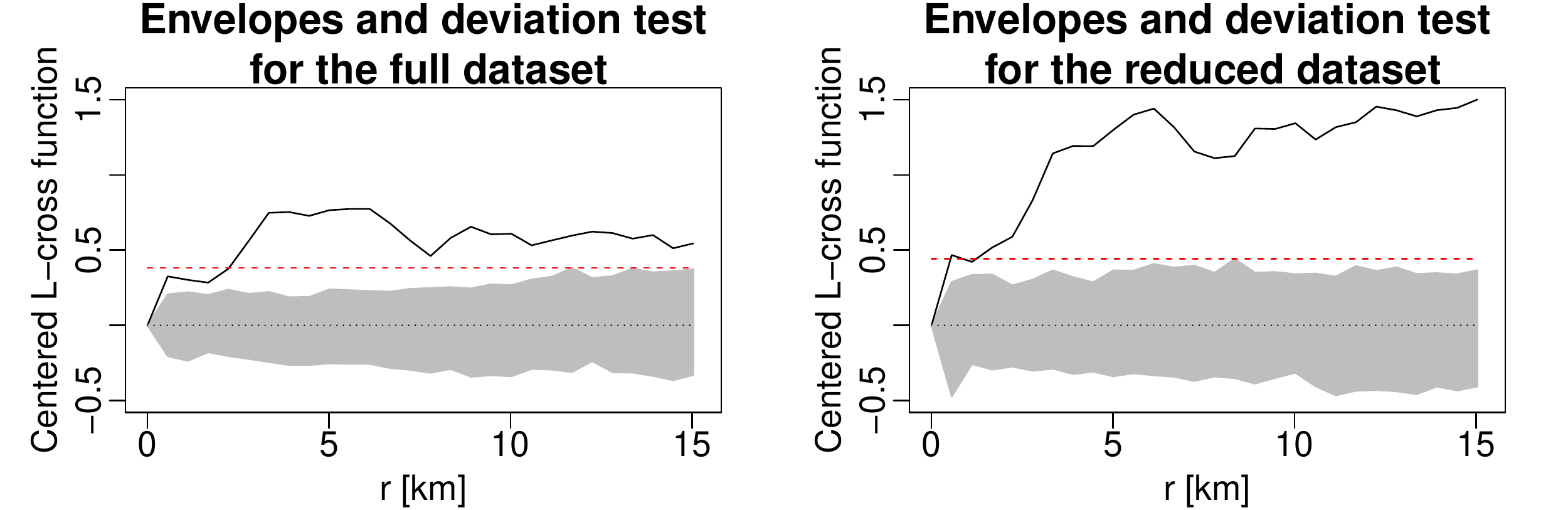}
\caption{Black solid lines represent $\hat{L}_{cross}(r)-\mathbb{E}[\hat{L}_{cross}(r)]$ for the observed pattern, the $95\%$ envelopes (gray areas) are obtained using 99 simulated patterns and the red dashed lines indicate the upper deviations. Deviation test: if the black solid line rises above the red dashed line then the interaction can be considered significant at significance level $\alpha=0.05$. The values of $\mathbb{E}[\hat{L}_{cross}(r)]$ are estimated using independently simulated point patterns generated according to the null hypothesis.}\label{fig:k_cross_env}
\end{figure}

The null hypothesis is rejected for both the full and the reduced dataset (see Figure \ref{fig:k_cross_env}). For the reduced dataset this provides evidence of a stronger clustering effect. The R code used to perform this test and produce Figure \ref{fig:k_cross_env} is given in \ref{supp:data_codes}. Application of the same deviation test on the bivariate $L$-functions $\hat{L}_{ij}(r)$ provides an indication of which couples of placenames exhibit significant interaction (see Figure \ref{plots-fig:interaction_graph} of \ref{supp:tables_plots}).

The preliminary analysis we just presented indicates a clustering interaction between points of different types.
Nevertheless $K$-functions do not provide explicit estimates and quantification of uncertainty for the parameters of interest (including the cluster partition itself).
In the next section we develop a more advanced model in order to provide more informative answers to the historians questions.
%$K$-functions can indicate interaction and suggest which pairs of placenames interact more than others. Nevertheless they do not provide explicit estimates.
We regard $K$-functions as a useful exploratory tool and the fact that they indicate interaction is a motivation to pursue further statistical analysis.
%to guide us in the data analysis and to double-check the results we obtain from our model.% neither of the global parameters we are interested in (e.g. $\sigma$ or the ratio of clustered points versus non-clustered), neither of the partition into clusters $\rho$.

We note that Dr.\ Stuart Brookes from UCL has already used second moment functions to do some preliminary analysis on the Anglo-Saxon settlements dataset presented here (personal communication by Prof.\ John Blair).

\section{A Bayesian complementary clustering model}\label{sec:AS_model}
%\section{The Anglo-Saxon settlements model}\label{sec:AS_model}

\subsection{Random Partition Models}
We present Random Partition models (RPMs) in the specific context of planar $k$-type point processes. For more general and detailed discussions see \citet{Lau2007} and \citet{Muller2010}. Let $\rho$ be a partition of an ordered set of marked points $\textbf{x}=\left((x_1,m_1),\dots ,(x_{n(\textbf{x})},m_{n(\textbf{x})})\right)$, with each $(x_i,m_i)$ belonging to $\R^2\times \{1,\dots ,k\}$. Thus $\rho$ can be represented as an \emph{unordered} collection $\{C_1,\dots ,C_{N(\rho)}\}$ of disjoint non-trivial subsets of the indices $\{1,\dots ,n(\textbf{x})\}$ whose union is the whole set $\{1,\dots ,n(\textbf{x})\}$. RPMs are used to draw inferences on the partition $\rho$ given the observed points \textbf{x}.
Given $C_j=\left\{i_1^{(j)},\dots ,i_{s_j}^{(j)}\right\}$ we define $\textbf{x}_{C_j}=\Big\{\big(x_{i_1^{(j)}},m_{i_1^{(j)}}\big),\dots ,\big(x_{i_{s_j}^{(j)}},m_{i_{s_j}^{(j)}}\big)\Big\}$, for $j$ running from 1 to $N(\rho)$.
We call $\textbf{x}_{C_j}$ cluster and $s_j$ the size of the cluster.
Given the partition $\rho$, we suppose that locations in each cluster $\textbf{x}_{C_j}$ are generated independently of locations in other clusters, according to a probability density function $h_{(s_j,\sigma)}(\cdot)$ depending on $s_j$ and on a global intra-cluster dispersion parameter $\sigma$. Thus the probability density function of \textbf{x} conditional on $\rho$ and $\sigma$ is $\prod_{j=1}^{N(\rho)}h_{(s_j,\sigma)}(\textbf{x}_{C_j})$.

We assign independent prior distributions to $\rho$ and $\sigma$. With a slight abuse of notation, we denote them by $\pi(\rho)$ and $\pi(\sigma)$ respectively. We require $\pi(\rho)$ to be exchangeable with respect to the point indices $\{1,\dots ,n(\textbf{x})\}$ to reflect the fact that point labels are purely arbitrary and have no specific meaning. We obtain the following expression for the posterior density function
\begin{equation*}
\pi(\rho,\sigma|\textbf{x})\quad\propto\quad\pi(\rho)\;\pi(\sigma)\prod_{j=1}^{N(\rho)}h_{(s_j,\sigma)}(\textbf{x}_{C_j})\,.
\end{equation*}

\subsection{Likelihood function}\label{sec:likelihood}
Given $\rho$ and $\sigma$, each cluster $\textbf{x}_{C_j}$ is constructed as follows. First an unobserved center point $z_j$ is sampled from the observation region $W\subseteq\R^2$ with probability density function $g(\cdot)$. Then the observed points $x_{i_1^{(j)}},\dots ,x_{i_{s_j}^{(j)}}$ are given by
\begin{equation}\label{eq:cluster_construction}
x_{i_l^{(j)}}=z_j+y_{i_l^{(j)}}\;, \qquad l=1,\dots ,s_j
\end{equation}
where $y_{i_l^{(j)}}$ is defined as $w_{i_l^{(j)}}-s_j^{-1}\sum_{l=1}^{s_j}w_{i_l^{(j)}}$ with $w_{i_1^{(j)}},\dots ,w_{i_{s_j}^{(j)}}$ being independent bivariate $\textrm{N}(0,\frac{\sigma^2}{\pi}\I_2)$ random vectors, where $\I_2$ is the $2\times 2$ identity matrix. The variance parametrization $\frac{\sigma^2}{\pi}$ is chosen so that $\sigma$ equals the expected distance between two points in the same cluster, independently of the value of $s_j$. In fact if $x_1$ and $x_2$ belong to the same cluster it holds
$$
\mathbb{E}\bigg[\sqrt{(x_1-x_2)^\top(x_1-x_2)}\bigg]
=
\mathbb{E}\bigg[\sqrt{(w_1-w_2)^\top(w_1-w_2)}\bigg]
=
\sqrt{\frac{\pi}{2}}\sqrt{\frac{2\sigma^2}{\pi}}=\sigma\,,
$$
where $a^\top a=\sum_{i=1}^2a_i^2$ for $a$ in $\mathbb{R}^2$, and 
we used the fact that the euclidean norm of a two dimensional $N(0,\eta^2\I_2)$ random vector follows the Rayleigh distribution and its mean equals $\sqrt{\frac{\pi}{2}}\eta$ for $\eta\geq0$.

Finally the marks $m_{i_1^{(j)}},\dots ,m_{i_{s_j}^{(j)}}$ are sampled uniformly from the set $\big\{\{m_1,\dots ,m_{s_j}\}\subseteq\{1,\dots ,k\}\; |\; m_{l_1}\neq m_{l_2}\hbox{ for }l_1\neq l_2\big\}$.

The resulting likelihood function is
\begin{equation}\label{eq:likelihood}
h_{(s_j,\sigma)}(\textbf{x}_{C_j})
\;=\;
\frac{g\left(\overline{x}_{C_j}\right)\prod_{l_1,l_2\in C_j;\, l_1\neq l_2}\1(m_{l_1}\neq m_{l_2})}
{{{k}\choose{s_j}} s_j\;(2\sigma^2)^{s_j-1}}
\exp\left(-\frac{\pi\delta_{C_j}^2}{2\sigma^2}\right),
\end{equation}
(Section \ref{calculations-sec:likelihood} of \ref{supp:calculations} gives calculations) where $\overline{x}$$_{C_j}$ is the Euclidean barycenter of $\textbf{x}_{C_j}$ and $\delta_{C_j}^2=\sum_{i\in C_j}\big(x_i-\overline{x}_{C_j}\big)^\top \big(x_i-\overline{x}_{C_j}\big)$.

Here we treat $g(\cdot)$ as a known function. For the purposes of data analysis we will replace $g$ with an estimate using Gaussian kernel smoothing (see for example Figure \ref{plots-fig:density_estimate} of \ref{supp:tables_plots}) with bandwidth chosen according to the cross-validation method \citep[p.115-118]{Diggle2003} and edge correction performed according to \citet{Diggle1985}. Note that this replacement commits us to the use of a data-driven prior.
%Finally, the window $W$ is defined as Great Britain (coastline obtained from $mapdata$ $R$ package \citealp{mapdata}) cropping the region where the density $g$ falls below a certain threshold, approximately at the borders between England-Scotland and England-Wales border (see \ref{supp:sensitivity} for more details).

\begin{rmk}\label{rmk:uniform_marks}
Given the heterogeneity in the number of settlements across different placenames, the assumption of the marks being sampled uniformly seems not to be very realistic.
In \ref{supp:sensitivity} we propose an empirical Bayes approach to include non-uniformity of marks in the model while keeping the computation feasible and we present inferences under that assumption.
Here we retain the uniform marks assumption for simplicity and because the two approaches produce similar inferences.
Moreover the inferences with the uniform marks assumption are more conservative (see \ref{supp:sensitivity}) and therefore preferable in this context.
\end{rmk}
\begin{rmk}
This model does not constrain $x_{i_l^{(j)}}=z_j+y_{i_l^{(j)}}$ to lie in the observation region $W$.
To make the model more realistic one could condition the distribution of $y_{i_l^{(j)}}$ in \eqref{eq:cluster_construction} on $z_j+y_{i_l^{(j)}}\in W$ (in a sort of edge-correction manner).
Nevertheless in our application the density function $g$ is not concentrated on the borders and the values of $\sigma$ are small (below 10 kilometers) compared to the size of $W$. Therefore most correction terms would be negligible.
Moreover computing a correction term for each center point $z_j$ would result in a consistent additional computational burden for each step of the Markov chain Monte Carlo (MCMC) algorithm in Section \ref{sec:MCMC}. Therefore we avoid such correction terms here.
Note that, since such correction terms would increase the probability of points being clustered, this approximation has a conservative effect.
\end{rmk}

\subsection{Prior distribution on $\sigma$}\label{sec:prior_sigma}
History and context suggest some considerations regarding the expected intra-cluster dispersion ($\sigma$ between $3$ and $10$km). For example, a basic consideration is that settlements of the same cluster needed to be at no more than a few hours walking distance, in order for the inhabitants of the settlements to interact administratively and politically. Nevertheless we prefer not to impose strong prior information on $\sigma$, as this gives us the opportunity to see whether our study of geographical location is in accordance with available contextual information. Thus we use a flat uniform prior for $\sigma$, as for example it is recommended in \citet[Sec. 7.1]{Gelman2006}
$$\sigma\;\sim\;\textrm{Unif}(0,\sigma_{max})\,.$$
We set $\sigma_{max}=50$km.
Given the historical context, such an upper bound for $\sigma$ constitutes a safe and conservative assumption.
We tested other values of $\sigma_{max}$, namely $20$ and $100$ km, and the inferences presented in Section \ref{sec:application} were not sensible to such changes, which is in accordance with \citet[Sec.2.2]{Gelman2006}.

\subsection{Prior distribution on $\rho$}\label{sec:prior_rho}
We need to model a partition made up of many small clusters. In fact each cluster can contain at most $k$ points (one for each color), and the historians expect most of the original clusters to have had fewer than 6 settlements. Common RPMs usually result in clusters with many data points each and therefore do not seem to be appropriate to our case (see for example Remark \ref{rmk:rho_prior2}).
We now define a prior distribution $\pi(\rho)$ designed for situations where each cluster can have at most $k$ points, with $k$ small compared to the number of points $n$.
%Since they produce very similar inferences we present and use only one of the two here, the Poisson model. In \ref{supp:sensitivity} we describe the second model, based on the Dirichlet-Multinomial distribution, and we use it to perform sensitivity analysis.
%Since they produce almost identical inferences we present only one of the two here,, and we leave the description and results of the other In \ref{supp:sensitivity} we describe an alternative model, based on the Dirichlet-Multinomial distribution, and we use it to perform sensitivity analysis.

\subsubsection{Poisson Model for $\pi(\rho)$}\label{sec:poisson_model}
The number of clusters $N(\rho)$ follows a Poisson distribution with mean $\lambda$ and each cluster size $s_j$ is sampled from $\{1,\dots k\}$ according to a probability distribution $\textbf{p}^{(c)}=(p_1^{(c)},\dots ,p_k^{(c)})$. Note that in such a model the (unobserved) point process of centers $\{z_1,\dots ,z_{N(\rho)}\}$ is a Poisson point process with intensity measure $\lambda\, g(\cdot)$ and the number of observed points need not equal $n$. Conditioning on observing $n$ points, the induced prior distribution on $\rho$ is $\pi(\rho|\lambda,\textbf{p}^{(c)})\propto\prod_{j=1}^{N(\rho)}\lambda p_{s_j}^{(c)}$. We assign the following conjugate priors to $\lambda$ and $\textbf{p}^{(c)}$
$$\lambda\;\sim\;\textrm{Gamma}(k_\lambda,\theta_\lambda)\,,\qquad\textbf{p}^{(c)}=(p_1^{(c)},\dots ,p_k^{(c)})\;\sim\;\textrm{Dir}(\alpha_1^{(c)},\dots ,\alpha_k^{(c)})\,.$$
Combinations of the following choices of hyperparameters did not change the posterior significantly: $k_{\lambda}=100,300,600$; $\theta_{\lambda}=0.5,1,3$ and $(\alpha_1^{(c)},\dots ,\alpha_k^{(c)})=(1/k,\dots ,1/k)$, $(1,\dots ,1)$ and $(1,1/(k-1)\dots ,1/(k-1))$. In the data analysis of Section \ref{sec:application} we set $k_{\lambda}=300$, $\theta_{\lambda}=1$ and $(\alpha_1^{(c)},\dots ,\alpha_k^{(c)})=(1/k,\dots ,1/k)$.

\begin{rmk}\label{rmk:rho_prior2}
In the RPMs literature it is common to assign a Dirichlet Process (DP) prior to $\rho$, $\pi(\rho\,|\,\theta)\propto\prod_{j=1}^{N(\rho)}\theta (s_j-1)!\,$, with concentration parameter $\theta$ either fixed or random. A DP prior (conditioning on having no cluster with more than $k$ points) would be equivalent to the Poisson model with fixed $\textbf{p}^{(c)}$ given by $p_l^{(c)}=\frac{(l-1)!}{\sum_{l=1}^k(l-1)!}$, for $l=1,\dots ,k$. Such a choice would enforce most clusters to have almost $k$ points and thus is not appropriate to this context where we expect most clusters to be smaller.
\end{rmk}

\begin{rmk}\label{rmk:dir_mult_model}\label{rmk:dir_model}
In \ref{supp:sensitivity} we describe an alternative model for $\pi(\rho)$, based on the Dirichlet-Multinomial distribution rather than the Poisson one.
Although the inferences we obtain from the two models are almost equivalent, the Poisson model is preferable because its posterior distribution factorizes over clusters and thus allow for cheaper computation.
\end{rmk}

\subsection{Model parameters and Posterior Distribution}
\begin{figure}[h]
\centering  \includegraphics[width=0.35\linewidth]{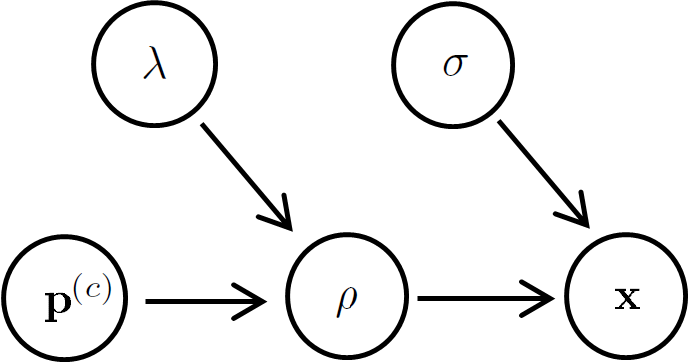}
  \caption{Conditional independence of the random elements involved in the Poisson Model.}\label{fig:cond_ind_graph}
\end{figure}

The model presented above results in the following unknown elements
\begin{equation*}
(\rho,\sigma,\textbf{p}^{(c)},\lambda)\;\in\;\Pn\times\R_+\times[0,1]^k\times\R_+,%\tag{Poisson}%\\
%(\rho,\sigma,\textbf{p})\;&\in\;\Pn\times\R_+\times[0,1]^k,\tag{Dirichlet-Multinomial}
\end{equation*}
where $\Pn$ is the set of all partitions of $\{1,\dots ,n\}$. Figure \ref{fig:cond_ind_graph} provides graphical representations of the underlying conditional independence structure.
Given the prior and likelihood distributions described in Sections \ref{sec:likelihood}, \ref{sec:prior_sigma} and \ref{sec:prior_rho} we obtain the following conditional posterior distributions
\begin{multline}
\pi(\rho\;|\;\textbf{x},\sigma,\textbf{p}^{(c)},\lambda)\propto\\
\prod_{j=1}^{N(\rho)}\left(\frac{g\left(\overline{x}_{C_j}\right)\lambda\, p^{(c)}_{s_j}}{c_{s_j}\sigma^{2(s_j-1)}}\;
\exp\left(-\frac{\pi\delta_{C_j}^2}{2\sigma^2}\right)\;\prod_{i,l\in C_j,\; i\neq l}\1(m_i\neq m_l)\right)\;,\label{eq:rho_poisson}
\end{multline}
\begin{align}
\pi(\sigma\;|\;\textbf{x},\rho,\textbf{p}^{(c)},\lambda)
\quad\propto&\quad
\frac{\1_{(0,\sigma_{max})}(\sigma)}{\sigma^{2(n-N(\rho))}}\,\exp
\left(\frac{\pi\sum_{j=1}^{N(\rho)}\delta_{C_j}^2}{2\sigma^2}\right)\,,\label{eq:sigma_poisson}\\
\textbf{p}^{(c)}\;|\;\textbf{x},\rho,\sigma,\lambda
\quad\sim&\quad
\textrm{Dir}\left(\alpha_1^{(c)}+N_1(\rho),\dots ,\alpha_k^{(c)}+N_k(\rho)\right)\,,\label{eq:p_poisson}\\
\lambda\;|\;\textbf{x},\rho,\sigma,\textbf{p}^{(c)}
\quad\sim&\quad
\textrm{Gamma}\left(k_\lambda+N(\rho)\;,\;\theta_\lambda/(\theta_\lambda+1)\right)\,,\label{eq:lambda_poisson}
\end{align}
where $c_s={{k}\choose{s_j}} s_j\,2^{s_j-1}$ and $\1_{(0,\sigma_{max})}(\cdot)$ is the indicator function of $(0,\sigma_{max})$.

\subsection{The posterior distribution of the partition $\rho$}\label{sec:intractability}

The posterior distribution $\pi(\rho|\textbf{x},\sigma,\textbf{p}^{(c)},\lambda)$ in \eqref{eq:rho_poisson} is intractable, meaning that we cannot obtain exact inferences from it and even performing approximate inferences is challenging. In fact the posterior sample space $\Pn$ is too large (of order between $n!$ and $n^n$) to perform brute force optimization or integration, and the complementary clustering condition makes it not easy to move in the state space.
To make these statements more precise we describe $\pi(\rho|\textbf{x},\sigma,\textbf{p}^{(c)},\lambda)$ in terms of hypergraphs and then we consider complexity theory results regarding its intractability.
For simplicity we will denote $\pi(\rho|\textbf{x},\sigma,\textbf{p}^{(c)},\lambda)$ by $\hat{\pi}(\rho)$.

Note that, although we have little hope of solving the problem in its general form (see Section \ref{sec:intract_summary}), Monte Carlo methods, for example, can still give satisfactory results in specific applications.% In Section \ref{sec:MCMC} we develop MCMC techniques to perform approximate inferences and use careful diagnostic techniques to monitor its convergence.

\subsubsection{Formulation of the model in terms of hypergraphs}\label{subsec:model_as_hypergraph}
\begin{figure}[h!]
\centering  \includegraphics[width=0.95\linewidth]{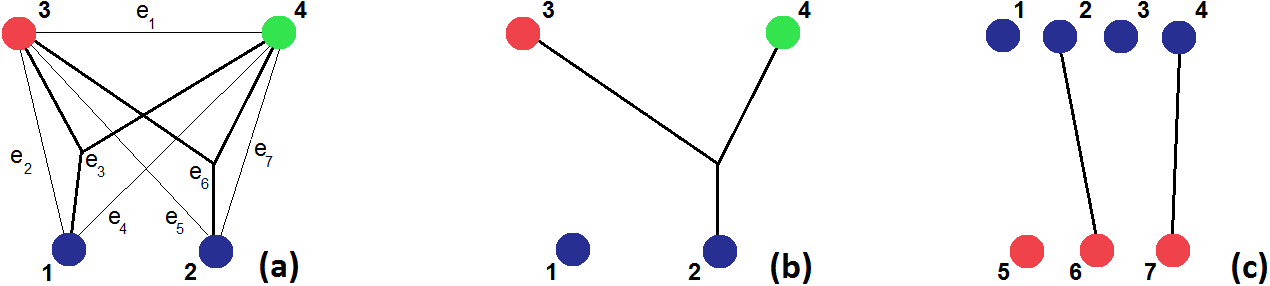}
  \caption{(a): Complete 3-partite hypergraph induced by the sets $V_1=\{1,2\}$, $V_2=\{3\}$ and $V_3=\{4\}$ corresponding to the colors blue, red and green. (b)-(c): Partial matching corresponding to $\rho=\big\{\{1\},\{2,3,4\}\big\}$ and $\rho=\big\{\{1\},\{2,6\},\{3\},\{4,7\},\{5\}\big\}$ respectively.}\label{fig:hypergraph_example}
\end{figure}

Hypergraphs are the generalization of graphs where each hyperedge can contain more than two vertices \citep{Berge1973}. In particular the complete $k$-partite hypergraph induced by $k$ sets $V_1,\dots,V_k$ is defined as $G=(V,E)$ where $V=V_1\cup\dots\cup V_k$ and $E=\{e\subseteq V\,:\,|e\cap V_l|\leq 1\;\forall\,l\,,\,|e|\geq 2\}$. See Figure \ref{fig:hypergraph_example} (a).
A partition $\rho\in\Pn$ of $n$ points into clusters is admissible for our model if and only if no cluster of $\rho$ contains two points of the same type.
Therefore a set of points is an admissible cluster if and only if the hyperedge connecting them belongs to the complete $k$-partite hypergraph induced by the $k$ set of points corresponding to the $k$ types.
Every admissible partition $\rho$ can then be interpreted as a partial matching (i.e. hypergraph with at most one hyperedge containing each point) contained in $G$ as follows: each cluster with at least two points corresponds to a hyperedge and each unlinked point is a cluster by itself (see Figure \ref{fig:hypergraph_example} (b)).
Moreover we can define a weight $w(e)$ for each hyperedge $e=\{x_1,\dots ,x_s\}$ in $E$,
\begin{equation}\label{eq:weightskd}
w (e)=
\frac{(c_1)^s\lambda\, p^{(c)}_{s}g\left(\overline{x}\right)\sigma^{-2(s-1)}}
{c_{s}\left(\lambda\, p^{(c)}_{1}\right)^s g(x_1)\cdots g(x_n)}
\exp\left(-\frac{\pi\sum_{i=1}^s\left(x_i-\overline{x}\right)^2}{2\sigma^2}\right),
\end{equation}
in such a way that $\hat{\pi}(\rho)$ is proportional to the weight of the matching $\rho$, defined as $\prod_{e\in \rho}w(e)$. In \eqref{eq:weightskd} $\overline{x}$ denotes the barycenter of $x_1,\dots ,x_s$.

In the remainder of the paper we will treat $\rho$ indifferently as a partition or as a matching, as the two formulations are equivalent. Note that in the two-color case $\rho$ reduces to a matching in a bipartite graph, see Figure \ref{fig:hypergraph_example} (c).

\subsubsection{Complexity theory results for $\hat{\pi}(\rho)$}\label{sec:intract_summary}
Given the hypergraph formulation of Section \ref{subsec:model_as_hypergraph} we can appeal to complexity theory results to obtain rigorous statements on the intractability of $\hat{\pi}(\rho)$.
In particular we consider the following tasks:
(a) finding the normalizing constant of $\hat{\pi}(\rho)$,
(b) finding the mode $\rho_{max}=\arg\max_{\rho\in\Pn}\hat{\pi}(\rho)$ and
(c) sampling from $\hat{\pi}(\rho)$.
In this section we briefly summarize the complexity of such tasks. \ref{supp:complexity} provides a more detailed analysis.
Note that the two-color case ($k=2$) and the multi-color case ($k\geq 3$) present substantially different complexity issues.% Motivated by the corresponding literature in the theory of algorithms we often refer to those as two-dimensional and k-dimensional case, even though in both cases our points lie on a plane.

\begin{enumerate}[(a)]
\item The normalizing constant of $\hat{\pi}(\rho)$ is the sum of the weights of all the matchings $\rho$ contained in $G$, that is the total weight of $G$. The problem of computing the total weight of a $k$-partite hypergraph is an $\#P$-hard counting problem \citep{Valiant1979}, even for $k=2$. The $\#P$-hard complexity class for counting problems is analogous to the \textit{NP}-hard complexity class for decision problems %. A counting problem $y$ is said to be $\#P$-hard if and only if every problem in $\#P$ (i.e.\ every polynomially checkable counting problem) is Cook-reducible to $y$ 
(see \citet{Valiant1979} or \citet{Jerrum2003} for definitions of these terms).
%The posterior mode $\rho_{max}$ can be efficiently computed for $k=2$, while it is an NP-hard problem for $k\geq 3$, although heuristic algorithms exist.

\item Finding the posterior mode $\rho_{max}=\arg\!\max_{\rho}\hat{\pi}(\rho)$ can be reduced to a $k$-dimensional optimal assignment problem (see \ref{supp:complexity}).
For $k=2$ this problem is efficiently solvable, for example in $O(n^3)$ steps with the Hungarian Algorithm \citep{Kuhn1955}. In contrast for $k\geq 3$ this is an NP-hard optimization problem.
Even more, unless P=NP, there is no deterministic polynomial-time approximation algorithm for a general cost function (i.e. the problem is not in $APX$). 
Heuristics algorithms exist, but no constant of approximation is provided (see \ref{supp:complexity}).
Therefore, while heuristics might still work in particular cases, the literature does not appear to provide a generic bounded-complexity method to obtain or approximate $\rho_{max}$.
%We might therefore resort to heuristics suited to our particular case.
%Therefore, assuming P$\neq$NP, we should not expect to perform exact posterior inferences in polynomial time.
%

\item For $k=2$, $\hat{\pi}(\rho)$ can be interpreted as a monomer-dimer system (see \ref{supp:complexity}).
\citet{Jerrum1996} describe a polynomial-time MCMC algorithm to draw approximate samples from $\hat{\pi}(\rho)$.
Unfortunately, the polynomial bound they provide on the number of MCMC steps needed is not practically feasible (more details in \ref{supp:complexity}). More recent results \citep{Karpinski2012} suggest that the techniques used by \citet{Jerrum1996} cannot be extended to $k\geq3$, and they prove a negative result for $k\geq6$ (see \ref{supp:complexity}).
\end{enumerate}

Theoretical results like the ones above do not rule out, for example, the possibility of obtaining approximate samples in specific situations, but do exclude the possibility of finding a scheme that does so (in polynomial time) for arbitrary instances of a certain class of distributions. Since the problem we consider is by no mean arbitrary it is feasible that special methods may produce good approximate samples. In Section \ref{sec:MCMC} we propose an MCMC algorithm for the two-color case and one for the $k$-color case. As a consequence of the results presented in this section it is clear that additional care is needed when empirically studying MCMC mixing properties.

\section{Description of proposed MCMC algorithm}\label{sec:MCMC}
We use the Metro\-po\-lis-within-Gibbs algorithm to sample from $\pi(\rho,\sigma,\textbf{p}^{(c)},\lambda|\textbf{x})$ given in \eqref{eq:rho_poisson}-\eqref{eq:lambda_poisson}.
Direct sampling from $\pi(\textbf{p}^{(c)}|\rho,\sigma,\lambda,\textbf{x})$ and $\pi(\lambda|\rho,\sigma,\textbf{p}^{(c)},\textbf{x})$ %given by \eqref{eq:p_poisson} and \eqref{eq:lambda_poisson} 
is straightforward and, given $(\rho,\textbf{p}^{(c)},\lambda,\textbf{x})$, few steps of the Metropolis-Hastings algorithm are sufficient for the distribution of $\sigma$ to be close to its stationary distribution $\pi(\sigma|\rho,\textbf{p}^{(c)},\lambda,\textbf{x})$.
In contrast sampling from $\pi(\rho|\textbf{x},\sigma,\textbf{p}^{(c)},\lambda)$, which for simplicity we will denote by $\hat{\pi}(\rho)$, is challenging (see Section \ref{sec:intract_summary}). To do this we use the Metropolis-Hastings (MH) algorithm. We consider ways of improving the efficiency and of assessing the convergence of MH algorithms in this framework.% For simplicity, in the remainder of Section \ref{sec:MCMC} we will denote $\pi(\rho|\textbf{x},\sigma,\textbf{p}^{(c)},\lambda)$ by $\hat{\pi}(\rho)$.

\subsection{2-color case}\label{sec:MCMC2d}
We commence by considering the two-color case because there is more known theory than in the general case and because the combinatorial structure of the sample space is simpler. We view $\rho$ as a matching in a bipartite graph with $n_1$ red points and $n_2$ blue points (see Section \ref{subsec:model_as_hypergraph}). We denote the edge connecting the $i$-th red point and the $j$-th blue point by the ordered couple $(i,j)\in\{1,\dots ,n_1\}\times\{1,\dots ,n_2\}$. 

The proposal $Q^{2D}(\rho_{old},\rho_{new})$ for $\rho$ is defined in two steps. First we select an edge $(i,j)$ according to some probability distribution $q_{\rho_{old}}(i,j)$ on $\{1,\dots ,n_1\}\times\{1,\dots ,n_2\}$. Then, having defined $i'$ as the index such that $(i',j)\in\rho_{old}$, if such an $i'$ exists, and similarly $j'$ as the index such that $(i,j')\in\rho_{old}$, if such a $j'$ exists, we propose a new state $\rho_{new}=\rho_{old}\circ (i,j)$ defined as
\begin{equation} \label{eq:ij_move}
\left\{
\begin{aligned}
%    \rho_{old}\circ (i,j)=\left\{%(\rho_{old},i,j)
%  \begin{array}
    \rho_{old}&+(i,j), &&\hbox{if neither }i'\hbox{ nor }j'\hbox{ exists,}& \hbox{(Addition)}\\
    \rho_{old}&-(i,j), &&\hbox{if }(i,j)\in\rho_{old}, & \hbox{(Deletion)}\\
    \rho_{old}&-(i,j')+(i,j), &&\hbox{if }j'\hbox{ exists and }i'\hbox{ does not exists,}& \hbox{(Switch)}\\
    \rho_{old}&-(i',j)+(i,j), &&\hbox{if }i'\hbox{ exists and }j'\hbox{ does not exists,} & \hbox{(Switch)}\\
    \rho_{old}&-(i',j)-(i,j') && &\\
    \ &+(i,j)+(i',j'), &&\hbox{if }i'\hbox{ and }j'\hbox{ exist and }(i,j)\notin\rho_{old},&\hbox{(Double-Switch)} 
%  \end{array}
\end{aligned}
\right.
\end{equation}
where $\rho-(i,j)$ and $\rho+(i,j)$ denote the matchings obtained from $\rho$ by respectively removing or adding the edge $(i,j)$. Display \eqref{eq:ij_move} defines the set of allowed moves starting from $\rho_{old}$ and it induces a neighbouring structure on the space of matchings as follows: $\rho_{new}$ is a neighbour of $\rho_{old}$ if $\rho_{new}=\rho_{old}\circ (i,j)$ for some $(i,j)$. \citet{Jerrum1996} and \citet{Oh2009} consider similar but slightly smaller sets of allowed moves, given by the addition and deletion moves and addition, deletion and switch moves, respectively. It is plausible that increasing the set of allowed moves improves the mixing of the MH Markov chain.

Display \eqref{eq:ij_move} does not identify uniquely the proposal $Q^{2D}(\rho_{old},\rho_{new})$ because we still need to choose $q_{\rho_{old}}(\cdot,\cdot)$. Different choices of $q_{\rho_{old}}(\cdot,\cdot)$ will affect the mixing properties of the MH algorithm. Previous works (e.g.\ \citet{Jerrum1996} and \citealp{Oh2009}) chose $q_{\rho_{old}}(i,j)$ to be a uniform measure over the edges $(i,j)\in E$. A naive implementation of such choice leads to poor mixing because most proposed matchings $\rho_{new}$ are improbable and therefore are typically rejected (in our experiments usually less than $1\%$ of the proposed moves were accepted). Some authors overcome this problem using a truncation approximation of the posterior: they force edge weights below a certain threshold $\delta$ to be zero, and then choose
\begin{equation}\label{eq:P1}
q_{\rho_{old}}(i,j)\;\propto\;\; \1_{\{w_{ij}>\delta\}}\;,\tag{P1}
\end{equation}
where $w_{ij}$ is the weight of the edge $(i,j)$ defined in \eqref{eq:weightskd} and $\1$ denotes the indicator function. See for example the measurement validation step in \citet{Oh2009}.

In the following we propose a choice of $q_{\rho_{old}}$ that achieves a better mixing than \eqref{eq:P1} and does so without requiring to target an approximation of the posterior.

Firstly note that, especially when $\hat{\pi}(\rho)$ has a factorization in terms of edge weights, it is straightforward to evaluate $\hat{\pi}$ up to a multiplicative constant on the set of neighbours of $\rho_{old}$ defined in \eqref{eq:ij_move}. For example, for the addition move, $\frac{\hat{\pi}(\rho_{old}\circ (i,j))}{\hat{\pi}(\rho_{old})}=w_{ij}$. Thus, one may be tempted to propose proportionally to $\hat{\pi}$ restricted on the set of allowed moves as follows
\begin{equation}\label{eq:P2}
q_{\rho_{old}}(i,j)\;\propto\;\;\hat{\pi}(\rho_{new})\qquad\hbox{where } \rho_{new}=\rho_{old}\circ (i,j)\,.\tag{P2}
\end{equation}
Such a choice, however, does not take into account the fact that the normalizing constants of $q_{\rho_{old}}(\cdot,\cdot)$ and $q_{\rho_{new}}(\cdot,\cdot)$ differ for $\rho_{old}\neq\rho_{new}$. As a consequence, for example, detailed balance conditions, $\frac{Q^{2D}(\rho_{old},\rho_{new})}{Q^{2D}(\rho_{new},\rho_{old})}=\frac{\hat{\pi}(\rho_{new})}{\hat{\pi}(\rho_{old})}$, are not satisfied, not even approximately. A better choice for $q_{\rho_{old}}(\cdot,\cdot)$ is
\begin{equation}\label{eq:P3}
 q_{\rho_{old}}(i,j)\propto\frac{\hat{\pi}\left(\rho_{new}\right)}{\hat{\pi}\left(\rho_{old}\right)+\hat{\pi}\left(\rho_{new}\right)},\qquad\hbox{where } \rho_{new}=\rho_{old}\circ (i,j).\tag{P3}
\end{equation}
Our experiments show that the latter choice leads to a significant improvement in the mixing of the MH Markov chain compared to \eqref{eq:P1} and \eqref{eq:P2} (see Section \ref{sec:conv_diag}). The main reason is that the MH algorithm induced by such proposal has a very high acceptance rate (usually above $99\%$) without changing the set of allowed moves. It can be shown that, under some regularity assumption on the weights, the proposal given by \eqref{eq:P3} satisfies detailed balance condition in the asymptotic regime (i.e.\ when the number of points tends to infinity), and this helps to explain why the acceptance rate is so high. Similarly one could also derive Peskun ordering arguments in the asymptotic regime. We omit those theoretical results here in favor of demonstrating the mixing improvement given by \eqref{eq:P3} using the convergence diagnostic techniques in Section \ref{sec:conv_diag}.

There is a trade-off between the complexity of the proposal and the mixing obtained (a complex proposal increases the cost of each step, while a poor proposal increases the number of MCMC steps needed). We seek a compromise with good mixing properties, like \eqref{eq:P3}, while still requiring little computation, like \eqref{eq:P1}. In Section \ref{calculations-sec:p4} of \ref{supp:calculations} we derive the following proposal distribution to try to obtain such goal
\begin{equation}\label{eq:P4}
q_{\rho_{old}}(i,j)\propto\left\{
\begin{aligned}
&q^{(add)}(i,j)&\hbox{if}(i,j)\notin\rho_{old},\\
&q^{(rem)}(i,j)&\hbox{if}(i,j)\in\rho_{old},
\end{aligned}
\right.\tag{P4}
\end{equation}
where $q^{(rem)}(i,j)=w_{ij}^{-1/2}$ and
\begin{multline*}
q^{(add)}(i,j)
\;=\;
\sqrt{w_{ij}}\,
\left(1-\sum_{j'\neq j}\frac{w_{ij'}-\sqrt{w_{ij'}}}{1+\sum_{s\neq i}w_{sj'}+\sum_{l}w_{il}}\right)\\
\left(1-\sum_{i'\neq i}\frac{w_{i'j}-\sqrt{w_{i'j}}}{1+\sum_{s\neq j}w_{i's}+\sum_{l}w_{lj}}\right)\;.
\end{multline*}
Note that $q^{(rem)}(i,j)$ and $q^{(add)}(i,j)$ do not depend on $\rho$ and can be precomputed at the beginning of the MCMC run. See Section \ref{sec:conv_diag} for discussion of performance.

\subsubsection{Scaling the proposal with a multiple proposal scheme}\label{sec:multiple_prop}
When using the MH algorithm on continuous sample spaces one can usually tune the variance of its proposal distribution to improve the efficiency of its algorithm (see for example \citealp{Roberts1997}). Given the very high acceptance rate obtained proposing according to \eqref{eq:P3} it is natural to consider the possibility of scaling our proposal in order to obtain longer-scale moves. The scaling problem for MH algorithms in discrete contexts has been considered, for example, in \citet{Roberts1998}. In that case the sample space was $\{0,1\}^N$, the vertices of the $N$-dimensional hypercube, and the scaling parameter, say $l$, was a positive integer representing the number of randomly-chosen bits to be flipped at any given proposal.

Unfortunately, because of the nature of our sample space, it is not so straightforward to scale the proposal distribution $Q^{2D}(\rho_{old},\rho_{new})$. One possibility is to scale by choosing $l$ edges, $\{(i_h,j_h)\}_{h=1}^l$, and performing $l$ moves defined in \eqref{eq:ij_move}, proposing $\rho_{new}=\rho_{old}\circ(i_1,j_1)\circ\cdots\circ(i_l,j_l)$. However the $l$ moves corresponding to $\{(i_h,j_h)\}_{h=1}^l$ cannot be performed independently: consider, for example, the case where $i_1=i_2$. We would then have to perform $l$ moves sequentially, at a computational cost being roughly $l$ times the one of a single move. Therefore scaling the proposal in such a way does not seem to be effective.

Instead, if the $l$ moves could be performed independently, it would be possible to implement a multiple proposal scheme using parallel computation, thus leading to a significant computational gain. This can be obtained by considering an approximation of our model, where points at a distance greater or equal than some $r_{max}$ have probability $0$ of being in the same cluster. The latter procedure is equivalent to the truncation procedure cited in Section \ref{sec:MCMC2d} and can be viewed as coming from the use of truncated Gaussian distributions to model points distribution within clusters, see \eqref{eq:cluster_construction}. Using this truncated model and diving the observed region into a grid, we defined a multiple proposal scheme where the $l$ moves are proposed and accepted/rejected simultaneously and independently. Therefore, at each MH step, such $l$ moves can be performed in an embarrassingly parallel fashion, meaning that they can be performed without the need for any communication between them. In \ref{supp:multiple_proposal} we give more details on the implementation and we show that in practice the mixing of the resulting MH algorithm improves by a factor roughly equal to $l$ itself (note that the maximum value of $l$ is bounded above, in a way that depends on $r_{max}$ and the size of the observation region $W$). A parallel-computing implementation of this algorithm would offer significant speed-ups (we anticipate speed ups by a factor around 8 for our dataset, see \ref{supp:multiple_proposal}). Such speed-ups would increase with the size of the dataset and window, making this proposal scheme especially relevant for applications to very large datasets. In \ref{supp:multiple_proposal} this scheme is presented and tested for fixed $\sigma$. In case $\sigma$ is varying, either one requires an upper bound on $\sigma$ or one needs different square grids for different values of $\sigma$.

\subsubsection{Convergence Diagnostics}\label{sec:conv_diag}
\begin{figure}[h!]
\centering  \includegraphics[width=\linewidth]{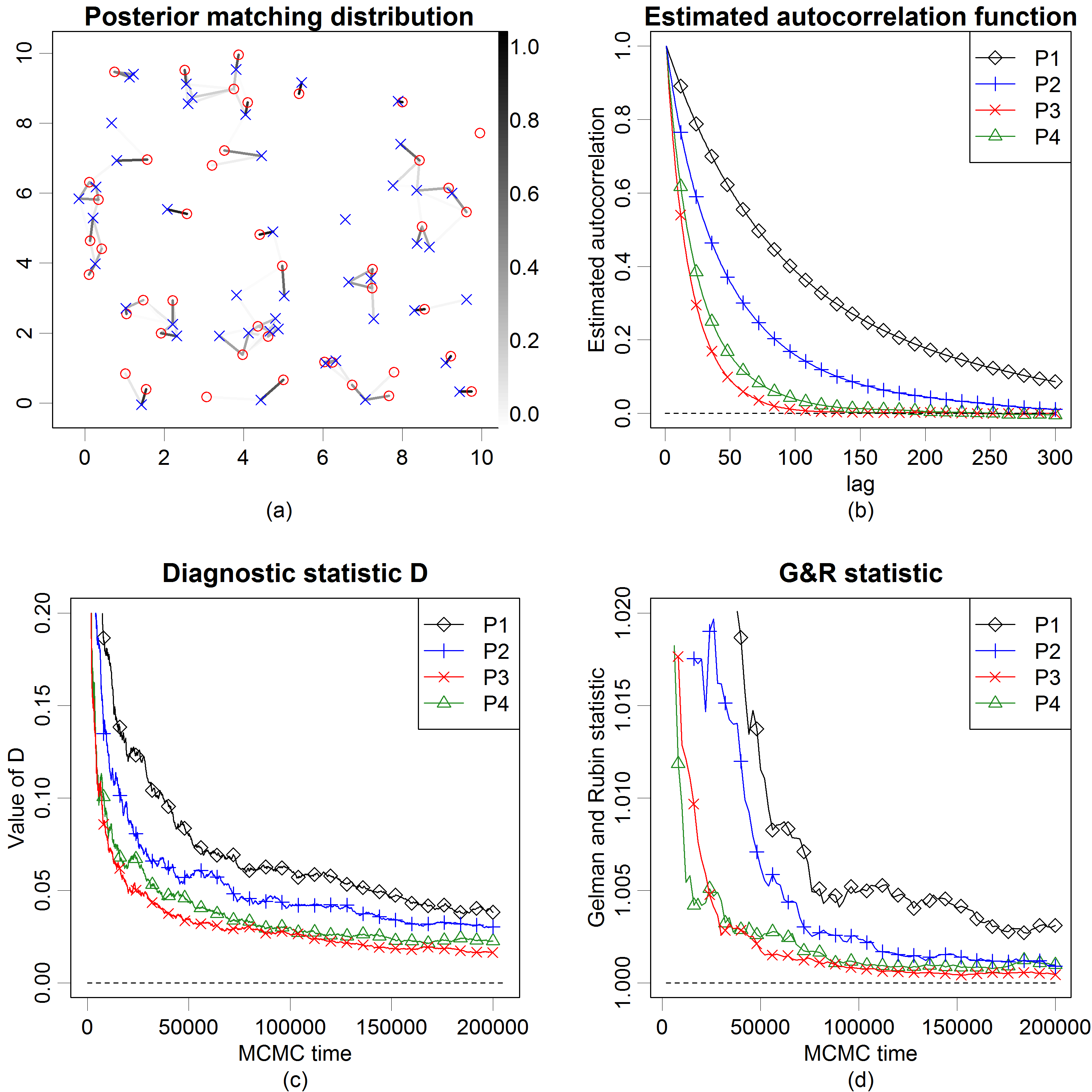}
 \caption{Four convergence diagnostic techniques described in Section \ref{sec:conv_diag}.}\label{fig:Conv_diag_output_an}
\end{figure}

We used various convergence diagnostic techniques in order to assess the reliability of our algorithm, to indicate the number of iterations needed, and to compare the efficiency of the four proposals \eqref{eq:P1}-\eqref{eq:P4} of Section \ref{sec:MCMC2d}. We demonstrate such techniques on the posterior $\pi(\rho|\sigma,\textbf{p}^{(c)},\lambda,\textbf{x})$ with $k=2$, $\sigma=0.3$, $p^{(c)}_{1}=p^{(c)}_{2}=0.5$, $\lambda=50$ and the center intensity $g(\cdot)$ being the uniform measure over $W=[0,10]\times[0,10]$. Here $\textbf{x}$ is a synthetic sample of $44$ red and $47$ blue points generated according to the model just defined, see Figure \ref{fig:Conv_diag_output_an} (a). We set the threshold $\delta$ of \eqref{eq:P1} to $0.001$. The R code used to produce the results presented in this Section is available in \ref{supp:data_codes}.

\begin{figure}[h!]
\centering  \includegraphics[width=\linewidth]{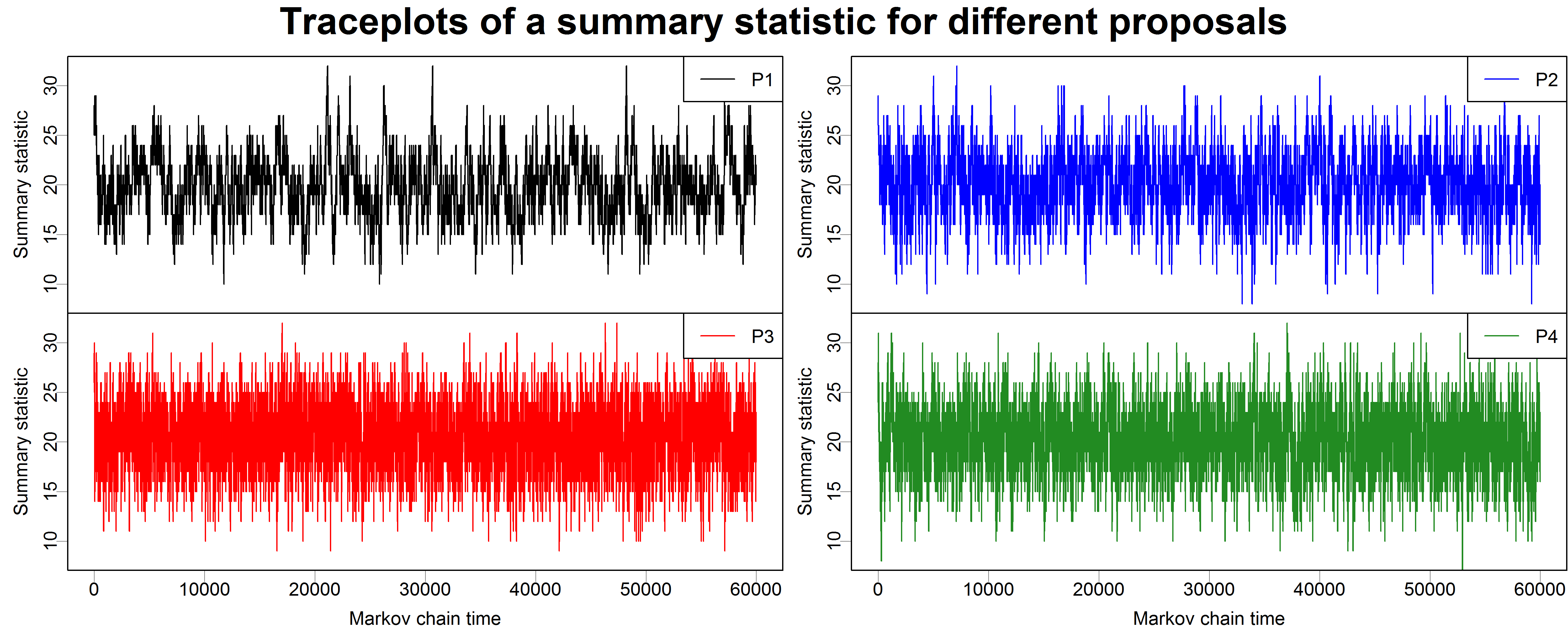}
 \caption{Traceplots of the number of differences from a reference matching.}\label{fig:traceplots}
\end{figure}

We first performed some qualitative output analysis by looking at summary plots of the MCMC samples of the partition (as the one in Figure \ref{fig:Conv_diag_output_an} (a)). Such plots can be helpful to spot when mixing has not yet occurred (see Section \ref{subsec:sim_temp}).

Secondly we considered different real valued summary statistics of the chain state (typically the number of different edges from some fixed reference matching). We plotted time series (see Figure \ref{fig:traceplots}) and empirical distributions of such real valued functions for different runs of the MCMC starting from different configurations. We estimated the autocorrelation functions (see Figure \ref{fig:Conv_diag_output_an} (b)), the Integrated Autocorrelation Time (IAT) and the Effective Sample Size (ESS) of such real-valued time series using the $R$ package $coda$ (see \citealp{Plummer2005}) in order to compare different versions of the algorithm (see Table \ref{table:compare_versions}).

Thirdly we used some standard convergence diagnostic techniques (see \citet{Brooks1998} and \citet{Cowles1996} for an overview of the techniques available). In particular we used the multivariate version of Gelman and Rubin diagnostic (see \citet{Gelman1992} and \citealp{BrooksGelman1998}). Figure \ref{fig:Conv_diag_output_an} (d) shows the results obtained by using a 10-dimensional summary statistic of $\rho$. In this context univariate summary statistics are not sufficiently informative and therefore misleading results can be obtained if these are used as the sole basis for convergence diagnostics.

Finally we compared two independent runs of the algorithm (with different starting states) by looking at estimates of the association probabilities $p_{ij}=\hbox{Pr}\big((i,j)\in\rho\big)$ with $\rho\sim\hat{\pi}$. We consider the following measure of proximity 
\begin{equation}\label{eq:proximity}
    D=\sup_{(i,j)\in E}|\hat{p}^{(1)}_{ij}-\hat{p}^{(2)}_{ij}|\,,
\end{equation}
where $\hat{p}^{(1)}_{ij}$ and $\hat{p}^{(2)}_{ij}$ denote the proportion of time that $(i,j)$ was present in the two MCMC runs. As starting states we considered the empty matching (each point is a cluster), the posterior mode (obtained with the Hungarian algorithm) and matchings obtained as the output of the MCMC itself. Since equation \eqref{eq:proximity} considers each link individually, we expect the resulting convergence diagnostic indicator $D$ to be more severe than the ones obtained from one or few summary statistics. Results are shown in Figure \ref{fig:Conv_diag_output_an} (d).

None of the convergence methods just presented indicate convergence issues except in the complete matching case (when the parameter $p^{(c)}_1$ is equal or very close to $0$), that is considered in the next subsection.
\begin{table}[h!]
\centering
\begin{tabular}{|c|c|c|c|c|c|c|c|}
  \hline
   & mean & Estimated & ESS for $10^4$ & steps [sec] & steps [sec] \\
   & acc.rate & IAT &  steps [for 1 sec]	& to $D<.05$ & to $GR <.005$ \\\hline
P1	&  17$\%$	&206	&262 [270]  	&1.4e05 [7.3]	&7.6e04	[13.5]  \\\hline
P2	&  41$\%$	&108	&544 [40]	&7.1e04	[84.6] 	&6.2e04	[97]  \\\hline
P3	&  97$\%$	&40		&1358 [99]	&2.0e04	[32.7] 	&2.4e04	[27.3]  \\\hline
P4	&  68$\%$	&55		&1038 [747]	&3.4e04	[2.2] 	&1.6e04	[4.8]  \\\hline
\end{tabular}\caption{Performances of the four proposals of Section \ref{sec:MCMC2d} on configuration in Figure \ref{fig:Conv_diag_output_an} (a) averaged over 5 independent runs for each proposal. $GR$ denotes the multivariate Gelman and Rubin statistic (potential scale reduction factor). The running time indicated in brackets is evaluated using R software on a desktop computer with Intel $i7$ processor.}\label{table:compare_versions}
\end{table}

All convergence diagnostic techniques agree in indicating that proposal \eqref{eq:P3} gives the best mixing; however in terms of real computation time the most efficient proposal is \eqref{eq:P4}. Note that such performances depend on the measure being targeted and, when running time is considered, on the computer implementation of such proposals. For the case considered in this Section, proposal \eqref{eq:P4} gives a 3-4 times speed-up over the commonly used choice \eqref{eq:P1}. Depending on the configuration such speed-up may vary. According to our experiments, for ``flatter'' distributions (e.g.\ increasing $\sigma$ to 1 and $p^{(c)}_1$ to 0.9, while keeping the other parameters unchanged) the speed-up almost disappears, while for ``rougher'' distributions (e.g.\ decreasing both $\sigma$ and $p^{(c)}_1$ to 0.1, while keeping the other parameters unchanged) the speed-up increases and \eqref{eq:P4} can be to 10 times faster than \eqref{eq:P1}.

\subsubsection{Multimodality and Simulated Tempering}\label{subsec:sim_temp}
\begin{figure}[h] \centering\includegraphics[scale=0.3]{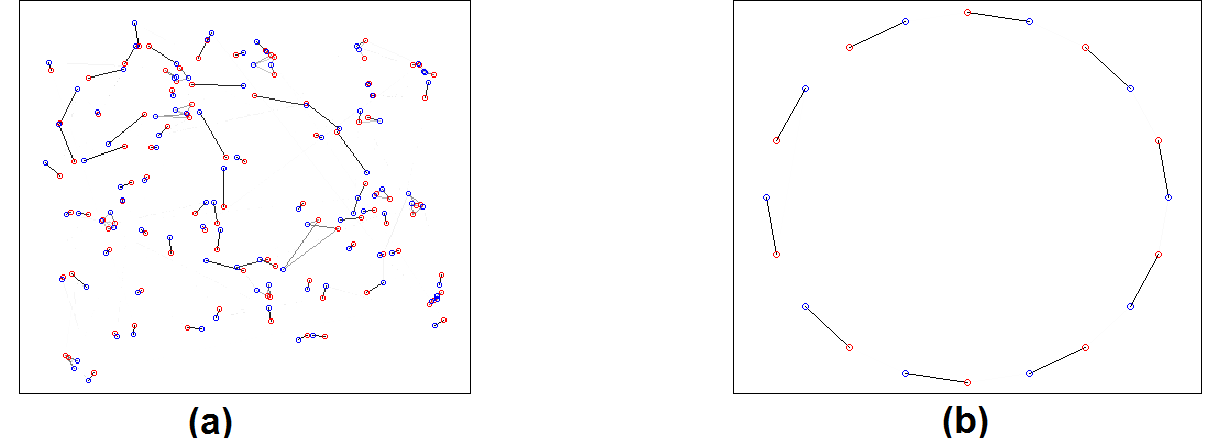}
\caption{Configurations corresponding to local maxima of $\pi(\rho|\textbf{x})$ for (a) a synthetic sample and (b) an artificially designed points configuration.}\label{fig:cycle1}
\end{figure}

In the complete matching case the posterior distribution of $\rho$ presents a strongly multimodal behavior. Cycle-like configurations like the one in Figure \ref{fig:cycle1} (a) are local maxima for $\hat{\pi}(\rho)$. In fact in order to reach a higher probability configuration (i.e.\ shorter links) from such a ``cycle'' configuration, with the set of allowed moves defined by \eqref{eq:ij_move}, the chain needs to pass through lower probability configurations (i.e.\ longer links). If we consider extreme cycle-like configurations (e.g.\ Figure \ref{fig:cycle1} (b)), then the MCMC will typically to get stuck in such local maxima.
In order to overcome this potential multimodality problem we implemented a simulated tempered version of our MCMC algorithm, see for example \citet{Geyer1995} or \citet{Marinari1992} for references. This technique manages to overcome local maxima for the complete matching case even when extreme cycle-like configurations are present (as in Figure \ref{fig:cycle1} (b)). Nevertheless our specific application do not present a complete matching case and therefore we have a milder multimodality and the MCMC algorithm exhibits sufficient mixing without the use of Simulated Tempering. Therefore Simulated Tempering is not used for the real data analysis, as convergence diagnostic tools do not show suspicious behavior.

We note that \citet{Dellaert2003} deal with multimodality in a similar posterior space (made of perfect matchings in a bipartite graph) arising from the Structure from Motion problem. In order to allow the MH algorithm to overcome local maxima like the one in Figure \ref{fig:cycle1} (b) they allow the MH proposal to include ``long'' moves that they call ``chain flipping''.

\subsection{k-color case}\label{sec:mcmc_kd}
We now define an MCMC algorithm that targets $\hat{\pi}(\rho)$ when $k\geq 3$. This case is harder than the two-dimensional one because it involves clusters with different dimensions and not just pairwise interaction.

\subsubsection{Description of proposed Gibbs projection MCMC algorithm}
\begin{figure}[h]
\centering  \includegraphics[width=\linewidth]{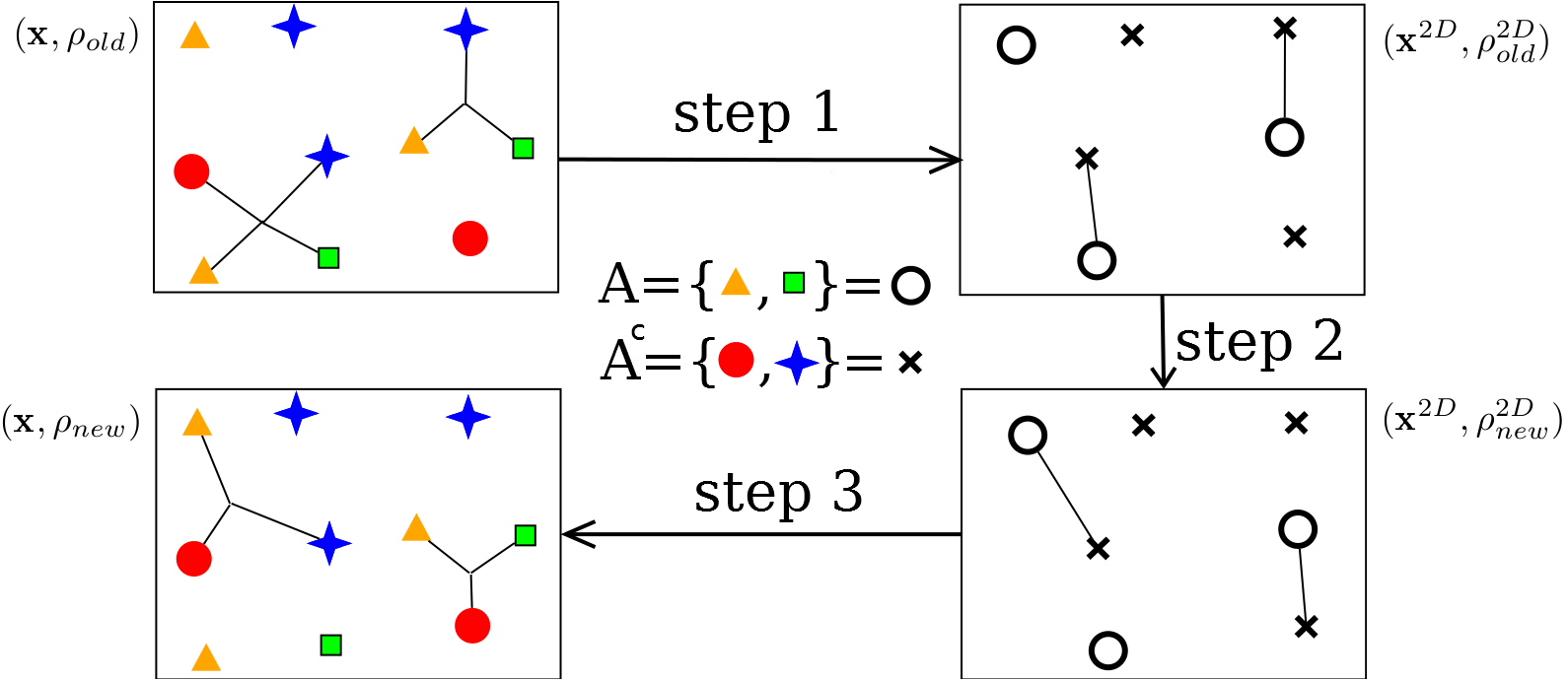}
  \caption{The action of a transition kernel $P^{(A)}$ for a given $A$.}\label{fig:kd_algorithm}
\end{figure}

We define the transition kernel $P$ of our MCMC algorithm as a mixture of ${k}\choose{\lfloor k/2 \rfloor}$ MH transition kernels, each of which corresponds to a group $A$ of $\lfloor k/2 \rfloor$ colors
\begin{equation}\label{eq:composition_of_proposals}
    P(\rho_{old},\rho_{new})= {{k}\choose{\lfloor k/2 \rfloor}}^{-1} \sum_{A\subset\{1,\dots ,k\},\,|A|=\lfloor k/2 \rfloor}P^{(A)}(\rho_{old},\rho_{new}),
\end{equation}
where $\lfloor k/2 \rfloor$ denotes the integer part of $k/2$ and ${k}\choose{\lfloor k/2 \rfloor}$ denotes a binomial coefficient. Here $P(\cdot,\cdot)$ selects a set of colors $A$, ``projects'' the $k$-color configuration to a 2-colors configuration where the new two colors correspond to $A$ and $A^c=\{1,\dots ,k\}\backslash A$ and then acts on the two-colors configuration. More precisely the action of $P^{(A)}$ is the following (see Figure \ref{fig:kd_algorithm}):
\begin{enumerate}
\item reduce the $k$-color configuration $(\textbf{x},\rho_{old})$ to a two-color one $(\textbf{x}^{2D},\rho^{2D}_{old})$ by replacing the points having colors in $A$ and $A^c$ respectively with their cluster centroids. We denote by $d_i$ the number of points merged together into the $i$-th point $x^{2D}_i$,
\item obtain $\rho^{2D}_{new}$ from $(\textbf{x}^{2D},\rho^{2D}_{old})$ with one or more MH moves using the proposal $Q^{2D}$ of Section \ref{sec:MCMC2d} on a target measure $\hat{\pi}^{2D}$ being the two-dimensional version of $\hat{\pi}$ (modified to take account of the multiplicity of the points $d_i$, see Section \ref{calculations-sec:correctness_k_dim} of \ref{supp:calculations}),
\item obtain the $k$-color configuration $(\textbf{x},\rho_{new})$ from $(\textbf{x}^{2D},\rho^{2D}_{new})$ by the inverse operation of Step 1 (note that here one needs to know what $A$ is).
\end{enumerate}
In order for this algorithm to be correct $\hat{\pi}^{2D}$ must be proportional to $\hat{\pi}$ on the collection of possible moves of $P^{(A)}$, so that $P^{(A)}$ satisfies detailed balance conditions with respect to $\hat{\pi}$. This follows from basic properties of the Gaussian density function and is proven in Section \ref{calculations-sec:correctness_k_dim} of \ref{supp:calculations}. Note that, when $k$ is even, $P^{(A)}$ is the same transition kernel as $P^{(A^c)}$. This is not an issue and it is indeed equivalent to never using $P^{(A^c)}$ and using $P^{(A)}$ twice more often.

By merging colors together we allow proposals which move many points at the same time from one cluster to another. Therefore the induced set of allowed moves is broader than, for example, the one of a scheme that moves one point at a time. \citet{Oh2009} consider also, for example, ``birth'' moves proposing to create a cluster from three or more single points in one step. Such moves are likely to be useful to speed up mixing in applications where there appear clusters with many points.

The mixture proposal in \eqref{eq:composition_of_proposals} allows us to re-use the two-color algorithm and in particular the approximation given in \eqref{eq:P4}.
In fact $\hat{\pi}^{2D}$ involves only pairwise interaction among points, meaning that $\hat{\pi}^{2D}(\rho^{2D})\propto\prod_{(i,j)\in\rho^{2D}}w^{2D}_{ij}$ for some weights $w^{2D}_{ij}$ depending on  $\textbf{x}^{2D}$ (see Remark \ref{calculations-rmk:pairwise} of \ref{supp:calculations}).
%involves only pairwise interaction among points (i.e.\ can be written as $\hat{\pi}^{2D}(\rho^{2D})\propto\prod_{(i,j)\in\rho^{2D}}w^{2D}_{ij}$)
%(see \eqref{calculations-eq:pi_2d_poisson} of \ref{supp:calculations}) is of the form
%$\hat{\pi}^{2D}(\rho^{2D})\propto\prod_{(i,j)\in\rho^{2D}}w^{2D}_{ij}$
Therefore, given $(\textbf{x}^{2D},\rho^{2D}_{old})$, it is possible to perform informed MH moves in the two-color matching space in a computationally efficient way using the approximation given in \eqref{eq:P4} (see Table \ref{table:compare_versions} for performances with two colors).
%For example, when using this algorithm for the real data analysis, each time we project on a two color configuration (Step 1 of Figure \ref{fig:kd_algorithm}), we performed 200 moves in the two color space (Step 2 of Figure \ref{fig:kd_algorithm}).
%   rather than in the original $k$-color one (note that the $\hat{\pi}^{2D}$ depends on $\textbf{x}$ only through $\textbf{x}^{2D}$).
%In particular the target measure $\hat{\pi}^{2D}$ (see \eqref{calculations-eq:pi_2d_poisson} of \ref{supp:calculations}) is of the form
%$\hat{\pi}^{2D}(\rho^{2D})\propto\prod_{(i,j)\in\rho^{2D}}w^{2D}_{ij}$

It would be desirable to design informed proposals like \eqref{eq:P3} or \eqref{eq:P4} directly in the $k$-color space, without the need of projecting on two-color subspaces.
However it would not be easy to do so in a computationally efficient way.
In fact, given the high-dimensionality of the space of matchings contained in a complete $k$-partite hypergraph, the set of neighbouring states $\rho_{new}$ of the current state $\rho_{old}$ would be extremely large.
Therefore it would be very expensive to use a scheme like \eqref{eq:P3} in this context.
Moreover, since $\hat{\pi}(\rho)$ involves interactions between three or more points, it would be difficult to design an approximation like \eqref{eq:P4} that can be evaluated efficiently.

Note that the mixture proposal in \eqref{eq:composition_of_proposals} first chooses uniformly at random a lower-dimensional subspace and then performs informed proposals in such a space.
Therefore such a scheme is a compromise between a ``fully uninformed'' proposal (which would choose uniformly at random some neighbour of $\rho_{old}$ and thus mix poorly) and a ``fully informed'' proposal (which, in order to make informed proposals in the $k$-color space, would be computationally expensive).
%One could still work directly in the $k$-color space using uninformed proposals (like choosing uniformly among the set of neighbouring states $\rho_{new}$

Since the $k$-color sample space is more complicated than the two-color one, additional care and longer MCMC runs are needed.
We implemented analogous convergence diagnostic techniques to the ones in Section \ref{sec:conv_diag}.
As might be expected, the number of MCMC steps needed to reach stationarity and to obtain mixing is much higher than in the two-color case (see end of Section \ref{sec:application}). Nevertheless our experiments suggest that, as in the two-color case, the MCMC manages to mix properly unless we are in a case close to complete matching (see Section \ref{subsec:sim_temp}).

\section{Analysis of Anglo-Saxon settlements with the Bayesian model}\label{sec:application}
%We now analyze the Anglo-Saxon settlements dataset supplied by Prof.\ John Blair, firstly by using common Spatial Statistic tools and then by applying the model described in Section \ref{sec:AS_model}. Such dataset is available in \ref{supp:data_codes}.
In this section we present the main results obtained by analyzing the Anglo-Saxon settlements dataset with the Random Partition Model described in Section \ref{sec:AS_model}.
The computation is done using the MCMC algorithm described in Section \ref{sec:MCMC}.
The analysis gives support to the historians hypothesis that settlements are clustered according to complementary functional placenames, and it permits inference about ranges of values for relevant parameters.

Here the no-clustering null hypothesis corresponds to $p^{(c)}_1=1$ (see Section \ref{sec:AS_model}).
\begin{figure}[h!]
\centering
\begin{minipage}[c]{.6\textwidth}
\centering
\includegraphics[width=\textwidth]{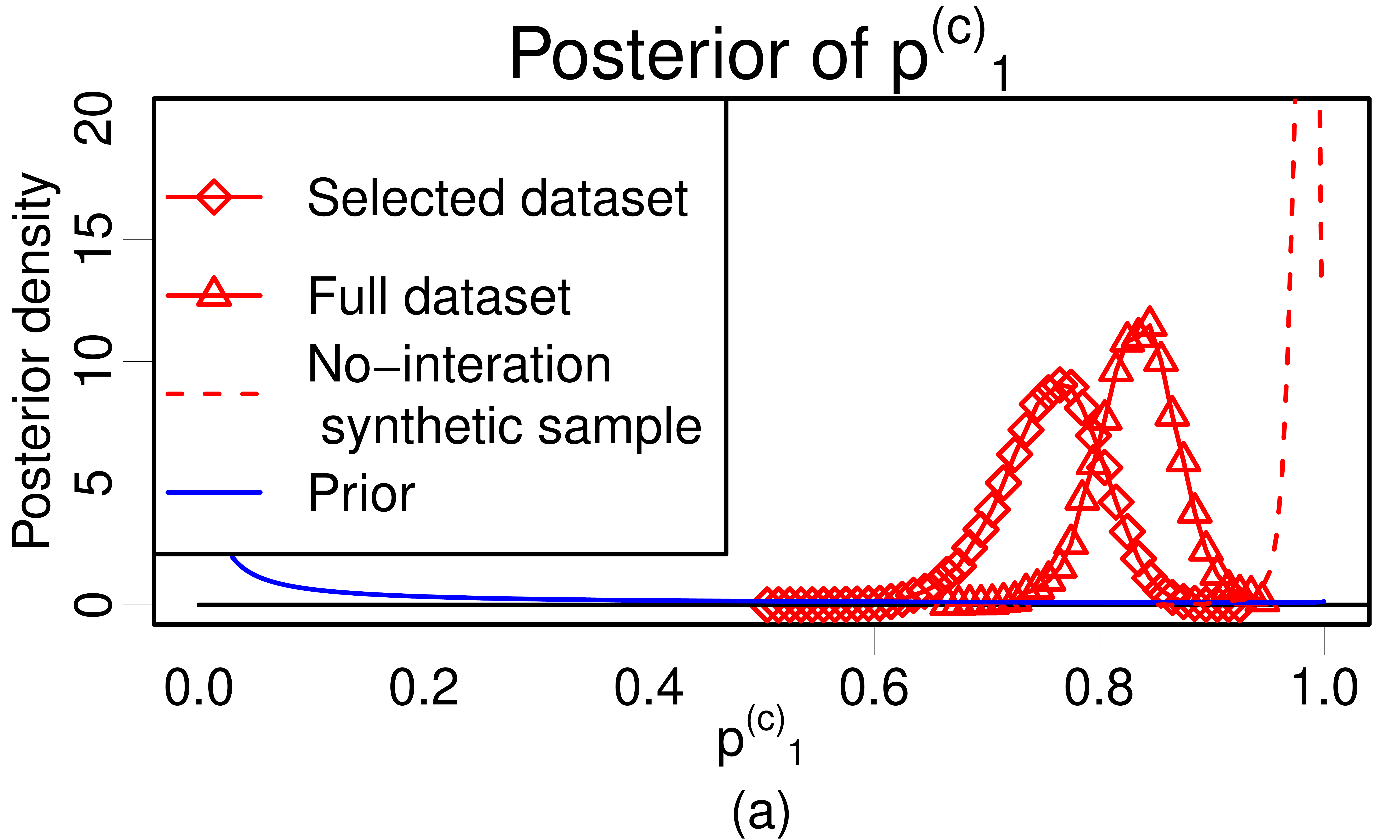}
\end{minipage}%
\hspace{.03\textwidth}%
\begin{minipage}[c]{.37\textwidth}
\centering
\includegraphics[width=\textwidth]{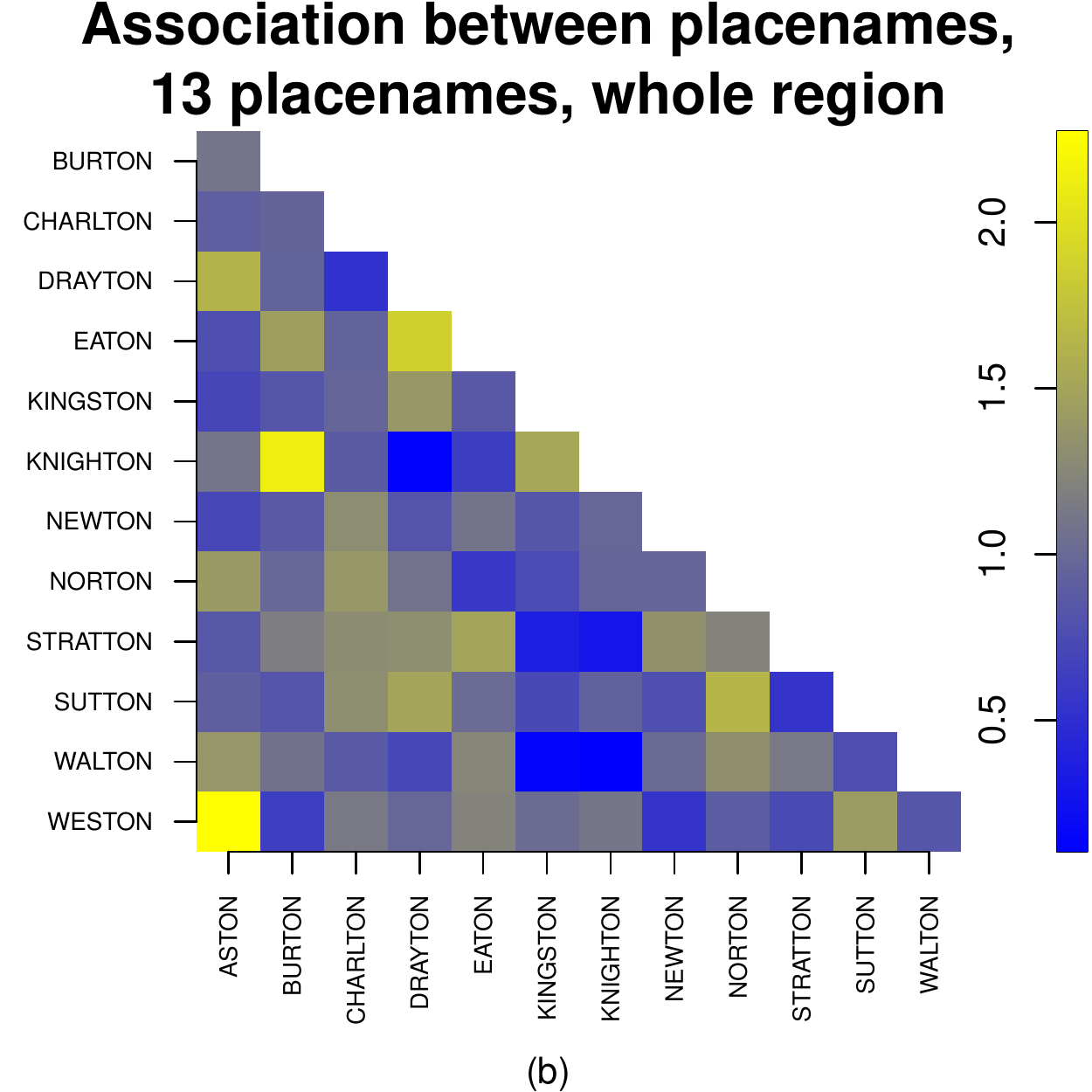}
\end{minipage}
\caption{(a) Estimated posterior distribution of $p^{(c)}_1$ (see Section \ref{sec:AS_model}) for the reduced and full dataset (13 and 20 placenames respectively). The no clustering hypothesis ($p^{(c)}_1=1$) lies outside the support of the posterior for the real dataset. (b) Measure of association between placenames (see end of Section \ref{sec:application}).}\label{fig:p_noise_ass}
\end{figure}
As shown in Figure \ref{fig:p_noise_ass}(a), such a hypothesis clearly lies outside the region where the posterior distribution is concentrated. %It can be clearly seen For both the reduced and full dataset analyzed with both models the no-clustering hypothesis is rejected by our statistical analysis (see Figure \ref{fig:p_noise_ass}).
As a sanity check we also fitted our model to synthetic samples generated according to the no-clustering null hypothesis of Section \ref{sec:null_hypothesis} (both with and without inhibition among points of the same type). As one would expect, in this case $p^{(c)}_1=1$ is included in the posterior support (see Figure \ref{fig:p_noise_ass}(a) for an example). 
%Figure \ref{fig:sigma_posterior} includes also the posterior distribution of the number of clustered points for a synthetic sample generated according to the no-clustering null hypothesis (dotted line). 

\begin{figure}[h!]
\centering
\includegraphics[width=\textwidth]{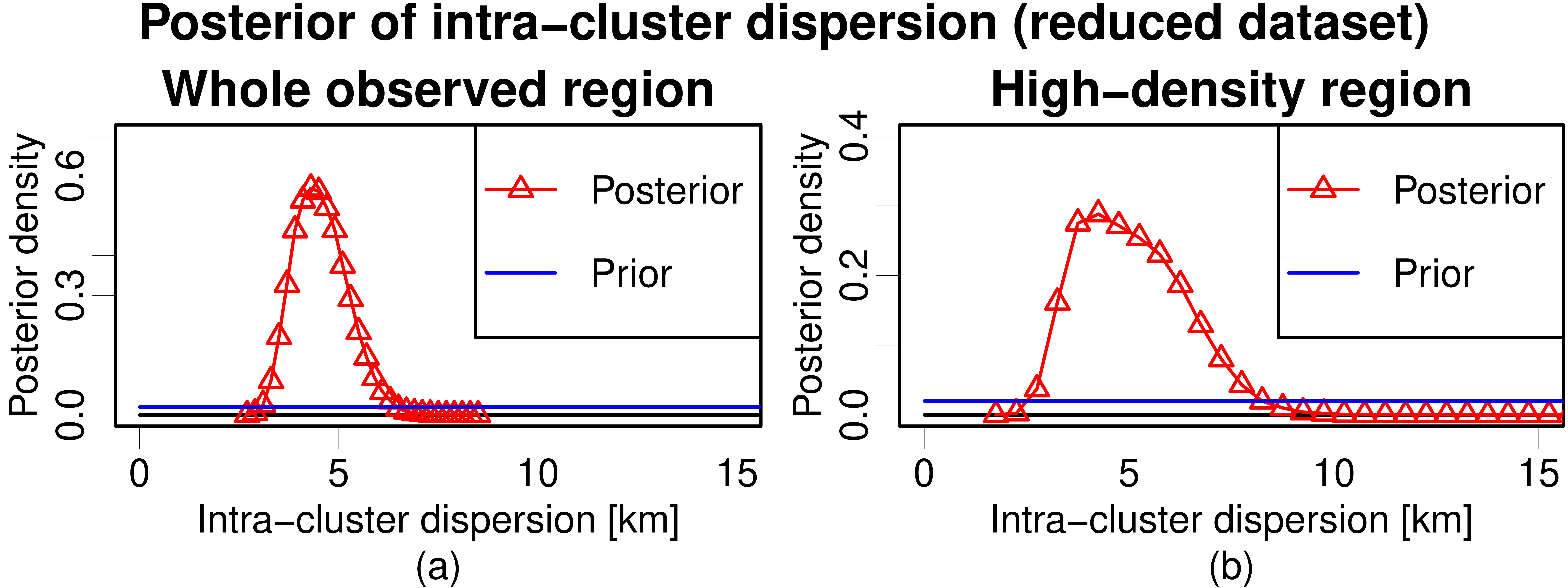}
\caption{(a) $\pi(\sigma|\textbf{x})$ for the reduced dataset. (b) $\pi(\sigma|\textbf{x})$ considering only a high-density region (see Section \ref{sec:discussion}).\label{fig:sigma_interaction}}
\end{figure}

Figure \ref{fig:sigma_interaction}(a) shows the estimated posterior distribution of $\sigma$ for the reduced dataset, which is clearly peaked around 4\,-\,5 km.
The $95\%$ Highest Posterior Density interval is $(3.3,5.9)$ km and the posterior mean is $4.6$ km. Therefore, according to the fit given by our model, the clustering behavior consists of clusters with settlements having distance being approximately $5$ km on average. It is satisfying to note that this value is in accordance with the value suggested by the historians involved in the project and coherent with the historical interpretation (see Section \ref{sec:prior_sigma}).
%Nevertheless, according to the posterior distribution of our models, the clustering behavior involves only half of the settlements approximately, meaning that only approximately half of the settlements belongs to clusters of size at least 2 (see left-hand side of Figure \ref{fig:sigma_posterior}).

\begin{figure}[h!]
\includegraphics[width=\textwidth]{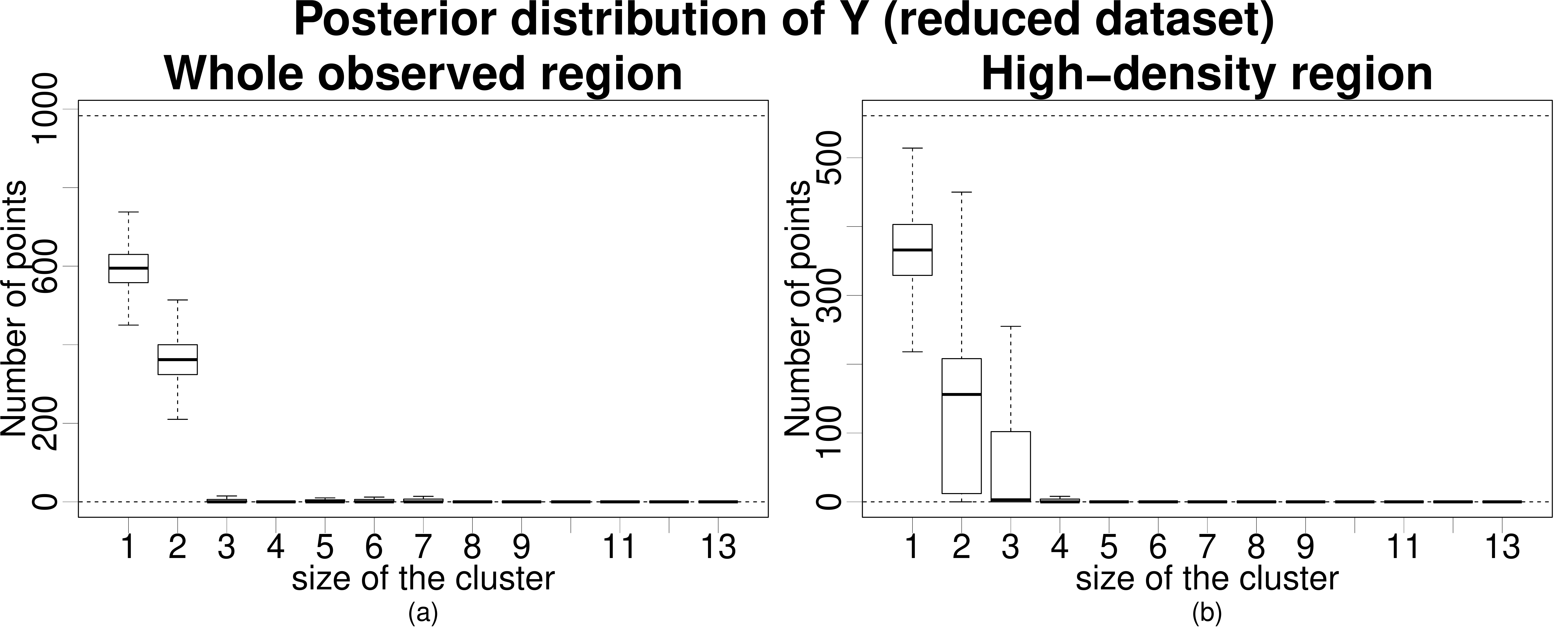}
\caption{(a) Posterior distribution of $\textbf{Y}=(Y_1,\dots,Y_k)$ for the reduced dataset. (b) Same but considering only the settlements in a high density region (see Section \ref{sec:discussion}).}\label{fig:Y_vec}
\end{figure}
Figure \ref{fig:Y_vec}(a) shows a box plot representation of the posterior distribution of $(Y_1,\dots ,Y_k)$, where $Y_l$ is the number of settlements in clusters of size $l$ (i.e.\ with $l$ settlements). Note that on average more than half of the settlements are not clustered (i.e\ they belong to clusters of size 1). Moreover most of the clustered settlements belongs to clusters of size 2. Historians expected to see more clusters involving three or four settlements than what was reported by our model. Inspection shows that model-fitting, and the requirement to fit clusters in the low-density region (which mostly contain couples with a high posterior probability), forces all the clusters in the high-density region to be couples too. In fact when the high-density region is analyzed separately (approximately 600 settlements) more triples appear and the posterior of $\sigma$ includes also slightly bigger values, see Figures \ref{fig:sigma_interaction}(b) and \ref{fig:Y_vec}(b). This suggests that there might be an heterogeneity in the clustering behaviour between high and low-density regions which is not captured in the model when applied to the whole region. This indicates a possible direction for future work (see Section \ref{sec:discussion}).

\begin{figure}[h!]
\includegraphics[width=\textwidth]{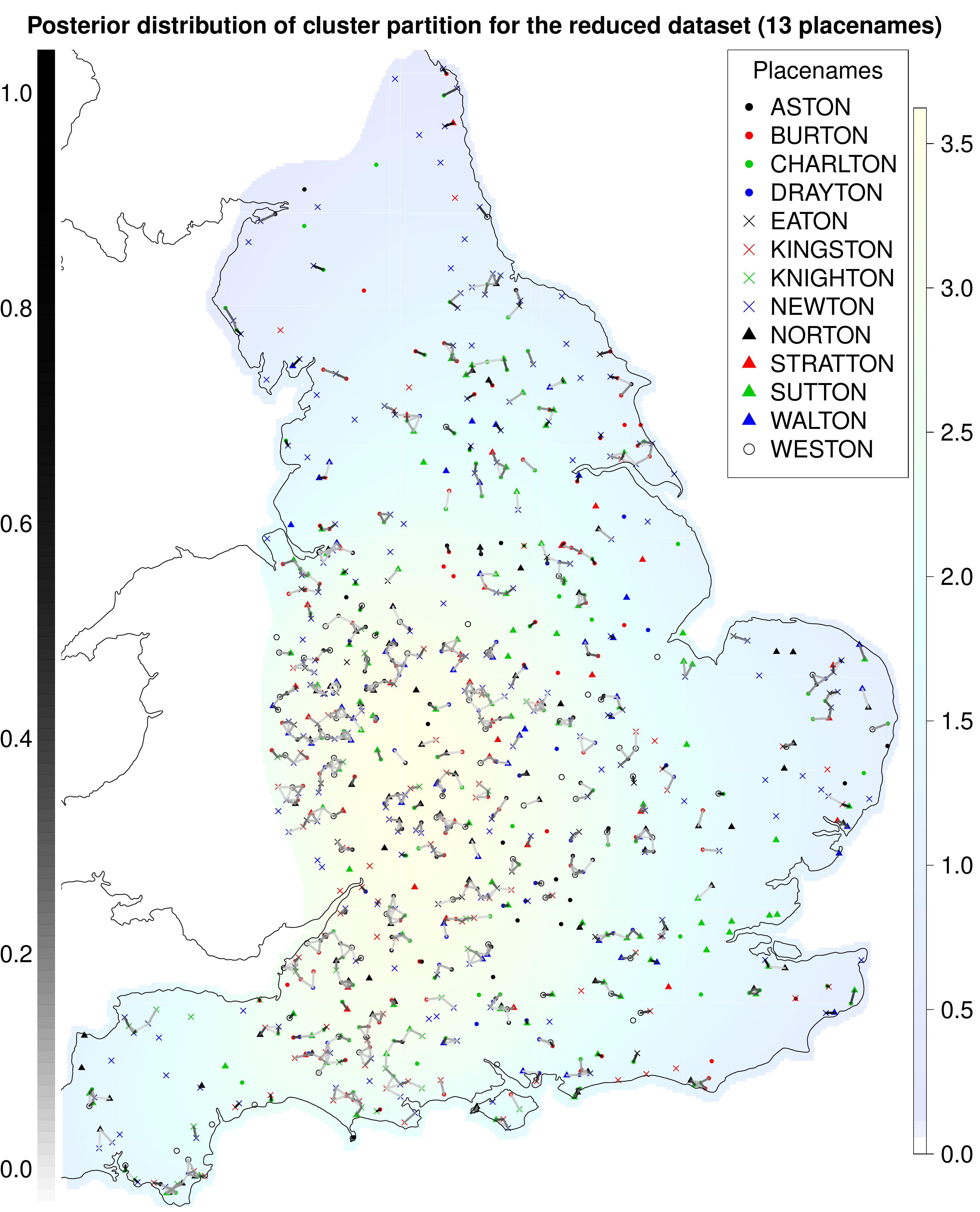}
\caption{Graphical representation of $\pi(\rho|\textbf{x})$, where $\textbf{x}$ is the reduced dataset (13 placenames) in the whole observed region. The intensity of gray corresponds to the estimated posterior probability of the cluster. The truncated kernel density estimation of $g$ is plotted in the background, with values express in relative terms with respect to the uniform measure.}\label{fig:grayplot}
\end{figure}
Figure \ref{fig:grayplot} shows a graphical representation of the posterior distribution of the partition $\rho$ for the reduced dataset. This representation is of considerable use since it provides a visual understanding of how the model is fitting the data and enables comparison with contextual information.

%Figure \ref{fig:sigma_interaction}(c) represents a measure of association between placenames. Given two placenames such measure is defined as the expected number of clusters including both placenames according to the posterior distribution $\pi(\rho|\textbf{x})$, divided by the expected number of such clusters according to a null hypothesis distribution. In the null hypothesis the partition $\rho$ follows the posterior distribution $\pi(\rho|\textbf{x})$ but, given $\rho$, the placename of each settlement is re-sampled (with probability of each placename proportional to its numerosity in the real data) independently of the others, conditioning on having no two settlements with the same placename in the same cluster. The expected values of interest under such distribution have been estimated using standard Monte Carlo methods. High values suggest positive interaction between placenames, while low values suggest the opposite. Most of the positive association suggested by Figure \ref{fig:sigma_interaction}(c), such as $Knighton-Burton$, $Eaton-Drayton$ or $Knighton-Kingston$, are coherent with the current historians hypothesis. 

%From the previous results it can be seen that the Poisson and the Dirichlet-Multinomial model lead to almost equivalent posterior distributions (e.g.\ Figures \ref{fig:sigma_interaction} and \ref{fig:Y_vec}). The models differ only in the way the prior distribution of the partition $\rho$ is built, so this appears to indicate insensitivity to the specification of this component of the prior. 
We perform sensitivity analysis on the values of the hyperparameters of $\sigma$, $\lambda$ and $\textbf{p}^{(c)}$ (see Section \ref{sec:AS_model} for details on tested values) and the posterior distribution did not seem to be much sensitive to their specification. As a further sensitivity analysis, in \ref{supp:sensitivity} we specify and implement an alternative model for the prior distribution of the partition $\rho$.

Figure \ref{fig:p_noise_ass}(b) represents a measure of association between placenames. Given two placenames, say $a$ and $b$, the measure is defined as 
\begin{equation}\label{eq:measure_of_association}
\frac{\Pr[A|B]}{\Pr[A]}\quad=\quad\frac{\Pr[A\cap B]}{\Pr[A]\cdot\Pr[B]}\quad=\quad\frac{\Pr[B|A]}{\Pr[B]},
\end{equation}
where $A$ and $B$ are the events of observing placename $a$ and $b$ respectively in a cluster chosen uniformly at random from the clusters of $\rho$, with $\rho$ distributed according to $\pi(\rho|\textbf{x})$.
In Figure \ref{fig:p_noise_ass}(b) we plot the value of \eqref{eq:measure_of_association}, estimated from the MCMC run, in relative terms with respect to a null hypothesis.
In the null hypothesis we first choose a cluster from $\rho$ as before and then, denoting the number of settlements in the cluster by $s$, we sample $s$ placenames independently of each other with placename probabilities proportional to their numerosity in the dataset, conditioning on having pairwise different placenames.
%the placename of each settlement is re-sampled (
%In the null hypothesis the cluster is chosen uniformly at random from the clusters of a partition $\rho$ distributed according to $\pi(\rho|\textbf{x})$ $\pi(\rho|\textbf{x})$ but, given $\rho$, the placename of each settlement is re-sampled (with probability of each placename proportional to its numerosity in the real data) independently of the others, conditioning on having no two settlements with the same placename in the same cluster.
The expected values of interest under the null distribution have been estimated using standard Monte Carlo methods. High values in Figure \ref{fig:p_noise_ass}(b) suggest positive interaction between placenames, while low values suggest negative interaction. Most of the positive associations suggested by Figure \ref{fig:p_noise_ass}(b), such as $Knighton$-$Burton$, $Weston$-$Aston$ or $Eaton$-$Drayton$, are coherent with the current historians hypothesis. We note that, for a fixed $\rho$, the measure in \eqref{eq:measure_of_association} reduces to the \emph{coefficient of association} used by ecologists to measure association between species \citep{Dice1945}. Many different measures of association have been proposed in the ecological literature (see e.g.\ \citealp{Janson1981}). We chose \eqref{eq:measure_of_association} because it is symmetric, clearly interpretable and our experiments suggest that \eqref{eq:measure_of_association} is not much influenced by the numerosity of placenames $a$ or $b$, unlike most measures proposed in \citet{Janson1981}.

In order to obtain the results presented in this section, the MCMC algorithm of Section \ref{sec:mcmc_kd} was run for $10^6$ steps, where at each step 200 moves of the two-color configuration $(\textbf{x}^{2D},\rho^{2D})$ were proposed. We assessed convergence using the methods described in Section \ref{sec:conv_diag} (e.g.\ the value of $D$ in \eqref{eq:proximity} was approximately $0.02$). 
The time needed for such runs using a basic $R$ implementation (available in \ref{supp:data_codes}) on a desktop computer with an Intel $i$-$7$ processor is approximately 40 hours.% when using the Poisson model.% and 160 hours when using the Dirichlet-Multinomial model (in the latter case the posterior of $\rho$ does not factorize and so more computation is needed at each MH step).

\section{Discussion}\label{sec:discussion}
We have designed a Random Partition Model (RPM) that is able to capture the clustering behaviour expected by the historians involved in the project.
With no strong prior information, the model produces estimates that are meaningful for the historical context and in accordance with contextual information (e.g.\ see the posterior distribution of $\sigma$ and the association between placenames in Figure \ref{fig:p_noise_ass}(b)). We also defined a flexible prior distribution for clusters partition that is designed for a ``small clusters'' framework (where each cluster has at most $k$ points with $k$ small). In doing so we developed a RPM to perform complementary clustering which is applicable to other contexts where one needs to find aggregations of elements of different types. For example Professor Susan Holmes from Stanford University suggests that, in biological contexts, species living in the same geographical area assemble by dissimilarity as they fill different ecological niches, resulting in clusters of complementary species.

We carefully considered the computational aspects of this problem. After considering related problems in the complexity theory literature (see Section \ref{sec:intractability}) we employed the Metropolis-Hastings (MH) algorithm. We proposed a choice of MH proposal distributions that, compared to the usual choices found in the literature, achieves a significantly better mixing by approximating detailed balance conditions (see Section \ref{sec:MCMC2d}). We developed a multiple proposal scheme to allow for parallel computation that could be relevant for applications to bigger datasets (see Section \ref{sec:multiple_prop}). Regarding convergence diagnostic we note that, when monitoring the convergence of the MCMC in the partition space, univariate summary statistics appear to be not sufficiently informative to be used as a basis for convergence diagnostics. Diagnostics based on multivariate summary statistics or on the matrix of the estimated association probabilities seem to give more robust results (see Section \ref{sec:conv_diag}).

Although the proposed model manages to capture the pattern we were looking for, there is much room for improvement. For example, a direction for future work could be to extend the model in order to capture the heterogeneity in the clustering behaviour between high and low-density regions (see Section \ref{sec:application}). One could try to do this by allowing the parameters $\textbf{p}^{(c)}$ and $\sigma$ to vary over different regions, maybe as a function of the points density, while taking care not to over-parametrize the model (the amount of data is limited).
An alternative approach would be to modify the metric we use to evaluate distances between settlements. For example one could use a non-euclidean distance, perhaps based on the inverse square root of the settlements density, in order to allow for larger clusters (meaning with points further apart) in less dense regions.
One could also try to model the dispersion of settlements in the same cluster with a non-Gaussian distribution having heavier tails. 

Another extension that could result in a better fit is to introduce spatial dependence of placenames probabilities. %relax the exchangeability assumption on the distribution of placenames.
In fact in our model, both under the assumption of uniform and non-uniform marks (see Remark \ref{rmk:uniform_marks}), the probability of choosing a certain placename does not depend on the location, while the data suggest that different placenames are more likely to be chosen in different regions.

The context suggests that we are observing a thinned version of the original settlements distribution. Nevertheless it is not obvious how to incorporate missing data in this model without making further assumptions that do not seem realistic and are not supported by the historical informations available (e.g. that in each cluster there is a settlement for each type).

An interesting direction for future work is to try to incorporate other sources of data in the model.  For example topographical information seem to be related to the settlements clustering (e.g.\ historians think that settlements named Burton are related to good viewpoints) it would be interesting to find an efficient way to incorporate them in the model.

\section*{Acknowledgments}Prof.\ Wilfrid Kendall for PhD supervision. Prof.\ John Blair for collaboration and arranging supply of data. EPSRC for funding through the CRiSM grant EP/D002060/1.

\begin{supplement}
\sname{Supplement A}\label{supp:calculations}
\stitle{Additional calculations and derivations}
\slink[url]{http://www2.warwick.ac.uk/gzanella/compclust\_supp\_a.pdf}
%\sdescription{}
\end{supplement}
\begin{supplement}
\sname{Supplement B}\label{supp:complexity}
\stitle{Computational complexity of the model}
\slink[url]{http://www2.warwick.ac.uk/gzanella/compclust\_supp\_b.pdf}
%\sdescription{}
\end{supplement}
\begin{supplement}
\sname{Supplement C}\label{supp:multiple_proposal}
\stitle{Multiple Proposal Scheme}
\slink[url]{http://www2.warwick.ac.uk/gzanella/compclust\_supp\_c.pdf}
%\sdescription{}
\end{supplement}
\begin{supplement}
\sname{Supplement D}\label{supp:sensitivity}
\stitle{Model extensions and variations}
\slink[url]{http://www2.warwick.ac.uk/gzanella/compclust\_supp\_d.pdf}
%\sdescription{}
\end{supplement}
\begin{supplement}
\sname{Supplement E}\label{supp:tables_plots}
\stitle{Additional plots}
\slink[url]{http://www2.warwick.ac.uk/gzanella/compclust\_supp\_e.pdf}
%\sdescription{}
\end{supplement}
\begin{supplement}
\sname{Supplement F}\label{supp:data_codes}
\stitle{Anglo-Saxon settlements dataset and R codes}
\slink[url]{http://www2.warwick.ac.uk/gzanella/compclust\_supp\_f.zip}
%\sdescription{}
\end{supplement}

\bibliography{SettlementsBibliography}

\begin{thebibliography}{36}
% BibTex style file: imsart-nameyear.bst, 2013-01-28
% Default style options (sort=1,type=nameyear).
% Used options (sort=1,type=nameyear).

\bibitem[\protect\citeauthoryear{Baddeley}{2010}]{Baddeley2010}
\begin{barticle}[author]
\bauthor{\bsnm{Baddeley},~\bfnm{Adrian}\binits{A.}}
(\byear{2010}).
\btitle{{Multivariate and marked point processes}}.
\bjournal{Handbook of spatial statistics}
\bpages{371--402}.
\end{barticle}
\endbibitem

\bibitem[\protect\citeauthoryear{Baddeley, Moller and
  Waagepetersen}{2000}]{Baddeley2000}
\begin{barticle}[author]
\bauthor{\bsnm{Baddeley},~\bfnm{AJ}\binits{A.}},
  \bauthor{\bsnm{Moller},~\bfnm{J}\binits{J.}} \AND
  \bauthor{\bsnm{Waagepetersen},~\bfnm{R}\binits{R.}}
(\byear{2000}).
\btitle{{Non- and semi-parametric estimation of interaction in inhomogeneous
  point patterns}}.
\bjournal{Statistica Neerlandica}
\bvolume{54}
\bpages{329--350}.
\end{barticle}
\endbibitem

\bibitem[\protect\citeauthoryear{Baddeley and Turner}{2005}]{spatstat}
\begin{barticle}[author]
\bauthor{\bsnm{Baddeley},~\bfnm{Adrian}\binits{A.}} \AND
  \bauthor{\bsnm{Turner},~\bfnm{Rolf}\binits{R.}}
(\byear{2005}).
\btitle{{Spatstat: an {R} package for analyzing spatial point patterns}}.
\bjournal{Journal of Statistical Software}
\bvolume{12}
\bpages{1--42}.
\bnote{{ISSN} 1548-7660}.
\end{barticle}
\endbibitem

\bibitem[\protect\citeauthoryear{Baddeley and
  Van~Lieshout}{1995}]{Baddeley1995}
\begin{barticle}[author]
\bauthor{\bsnm{Baddeley},~\bfnm{AJ}\binits{A.}} \AND
  \bauthor{\bsnm{Van~Lieshout},~\bfnm{MNM}\binits{M.}}
(\byear{1995}).
\btitle{{Area-interaction point processes}}.
\bjournal{Annals of the Institute of Statistical Mathematics}
\bvolume{47}
\bpages{601--619}.
\end{barticle}
\endbibitem

\bibitem[\protect\citeauthoryear{Becker, Wilks and Brownrigg}{2013}]{mapdata}
\begin{bmanual}[author]
\bauthor{\bsnm{Becker},~\bfnm{Richard~A}\binits{R.~A.}},
  \bauthor{\bsnm{Wilks},~\bfnm{Allan~R}\binits{A.~R.}} \AND
  \bauthor{\bsnm{Brownrigg},~\bfnm{Ray}\binits{R.}}
(\byear{2013}).
\btitle{{Mapdata: Extra Map Databases}}
\bnote{R package version 2.2-2}.
\end{bmanual}
\endbibitem

\bibitem[\protect\citeauthoryear{Berge and Minieka}{1973}]{Berge1973}
\begin{bbook}[author]
\bauthor{\bsnm{Berge},~\bfnm{C}\binits{C.}} \AND
  \bauthor{\bsnm{Minieka},~\bfnm{E}\binits{E.}}
(\byear{1973}).
\btitle{{Graphs and hypergraphs}}.
\bpublisher{Amsterdam: North-Holland publishing company}.
\end{bbook}
\endbibitem

\bibitem[\protect\citeauthoryear{Brooks and Gelman}{1998}]{BrooksGelman1998}
\begin{barticle}[author]
\bauthor{\bsnm{Brooks},~\bfnm{SP}\binits{S.}} \AND
  \bauthor{\bsnm{Gelman},~\bfnm{A}\binits{A.}}
(\byear{1998}).
\btitle{{General methods for monitoring convergence of iterative simulations}}.
\bjournal{Journal of Computational and Graphical Statistics}
\bvolume{7}
\bpages{434--455}.
\end{barticle}
\endbibitem

\bibitem[\protect\citeauthoryear{Brooks and Roberts}{1998}]{Brooks1998}
\begin{barticle}[author]
\bauthor{\bsnm{Brooks},~\bfnm{SP}\binits{S.}} \AND
  \bauthor{\bsnm{Roberts},~\bfnm{GO}\binits{G.}}
(\byear{1998}).
\btitle{{Assessing convergence of Markov chain Monte Carlo algorithms}}.
\bjournal{Statistics and Computing}
\bvolume{8}
\bpages{319--335}.
\end{barticle}
\endbibitem

\bibitem[\protect\citeauthoryear{Chiu et~al.}{2013}]{Stoyan1987}
\begin{bbook}[author]
\bauthor{\bsnm{Chiu},~\bfnm{SN}\binits{S.}},
  \bauthor{\bsnm{Stoyan},~\bfnm{D}\binits{D.}},
  \bauthor{\bsnm{Kendall},~\bfnm{WS}\binits{W.}} \AND
  \bauthor{\bsnm{Mecke},~\bfnm{J}\binits{J.}}
(\byear{2013}).
\btitle{{Stochastic geometry and its applications}}.
\bpublisher{John Wiley \& Sons}.
\end{bbook}
\endbibitem

\bibitem[\protect\citeauthoryear{Cowles and Carlin}{1996}]{Cowles1996}
\begin{barticle}[author]
\bauthor{\bsnm{Cowles},~\bfnm{MK}\binits{M.}} \AND
  \bauthor{\bsnm{Carlin},~\bfnm{BP}\binits{B.}}
(\byear{1996}).
\btitle{{Markov chain Monte Carlo convergence diagnostics: a comparative
  review}}.
\bjournal{Journal of the American Statistical Association}
\bvolume{91}
\bpages{883--904}.
\end{barticle}
\endbibitem

\bibitem[\protect\citeauthoryear{Dellaert et~al.}{2003}]{Dellaert2003}
\begin{barticle}[author]
\bauthor{\bsnm{Dellaert},~\bfnm{Frank}\binits{F.}},
  \bauthor{\bsnm{Seitz},~\bfnm{Steven~M}\binits{S.~M.}},
  \bauthor{\bsnm{Thorpe},~\bfnm{Charles~E}\binits{C.~E.}} \AND
  \bauthor{\bsnm{Thrun},~\bfnm{Sebastian}\binits{S.}}
(\byear{2003}).
\btitle{{EM, MCMC, and chain flipping for structure from motion with unknown
  correspondence}}.
\bjournal{Machine Learning}
\bvolume{50}
\bpages{45--71}.
\end{barticle}
\endbibitem

\bibitem[\protect\citeauthoryear{Dice}{1945}]{Dice1945}
\begin{barticle}[author]
\bauthor{\bsnm{Dice},~\bfnm{Lee~R}\binits{L.~R.}}
(\byear{1945}).
\btitle{Measures of the amount of ecologic association between species}.
\bjournal{Ecology}
\bvolume{26}
\bpages{297--302}.
\end{barticle}
\endbibitem

\bibitem[\protect\citeauthoryear{Diggle}{1985}]{Diggle1985}
\begin{barticle}[author]
\bauthor{\bsnm{Diggle},~\bfnm{Peter}\binits{P.}}
(\byear{1985}).
\btitle{A kernel method for smoothing point process data}.
\bjournal{Applied statistics}
\bpages{138--147}.
\end{barticle}
\endbibitem

\bibitem[\protect\citeauthoryear{Diggle}{2003}]{Diggle2003}
\begin{bbook}[author]
\bauthor{\bsnm{Diggle},~\bfnm{Peter~J}\binits{P.~J.}}
(\byear{2003}).
\btitle{{Statistical analysis of spatial point patterns}}.
\bpublisher{Edward Arnold}.
\end{bbook}
\endbibitem

\bibitem[\protect\citeauthoryear{Diggle, Eglen and Troy}{2006}]{Diggle2006}
\begin{bincollection}[author]
\bauthor{\bsnm{Diggle},~\bfnm{PJ}\binits{P.}},
  \bauthor{\bsnm{Eglen},~\bfnm{SJ}\binits{S.}} \AND
  \bauthor{\bsnm{Troy},~\bfnm{JB}\binits{J.}}
(\byear{2006}).
\btitle{{Modelling the bivariate spatial distribution of amacrine cells}}.
In \bbooktitle{Case Studies in Spatial Point Process Modeling}
\bpages{215--233}.
\bpublisher{Springer}.
\end{bincollection}
\endbibitem

\bibitem[\protect\citeauthoryear{Gelling and Cole}{2000}]{Gelling2000}
\begin{bbook}[author]
\bauthor{\bsnm{Gelling},~\bfnm{Margaret}\binits{M.}} \AND
  \bauthor{\bsnm{Cole},~\bfnm{Ann}\binits{A.}}
(\byear{2000}).
\btitle{{The landscape of place-names}}.
\bpublisher{Shaun Tyas}.
\end{bbook}
\endbibitem

\bibitem[\protect\citeauthoryear{Gelman}{2006}]{Gelman2006}
\begin{barticle}[author]
\bauthor{\bsnm{Gelman},~\bfnm{Andrew}\binits{A.}}
(\byear{2006}).
\btitle{Prior distributions for variance parameters in hierarchical models}.
\bjournal{Bayesian analysis}
\bvolume{1}
\bpages{515--534}.
\end{barticle}
\endbibitem

\bibitem[\protect\citeauthoryear{Gelman and Rubin}{1992}]{Gelman1992}
\begin{barticle}[author]
\bauthor{\bsnm{Gelman},~\bfnm{A}\binits{A.}} \AND
  \bauthor{\bsnm{Rubin},~\bfnm{D}\binits{D.}}
(\byear{1992}).
\btitle{{Inference from Iterative Simulation using Multiple Sequences}}.
\bjournal{Statistical Science}
\bvolume{4}
\bpages{457--511}.
\end{barticle}
\endbibitem

\bibitem[\protect\citeauthoryear{Geyer and Thompson}{1995}]{Geyer1995}
\begin{barticle}[author]
\bauthor{\bsnm{Geyer},~\bfnm{CJ}\binits{C.}} \AND
  \bauthor{\bsnm{Thompson},~\bfnm{EA}\binits{E.}}
(\byear{1995}).
\btitle{{Annealing Markov chain Monte Carlo with applications to ancestral
  inference}}.
\bjournal{Journal of the American Statistical Association}
\bvolume{90}
\bpages{909--920}.
\end{barticle}
\endbibitem

\bibitem[\protect\citeauthoryear{Grabarnik, Myllym{\"a}ki and
  Stoyan}{2011}]{Grabarnik2011}
\begin{barticle}[author]
\bauthor{\bsnm{Grabarnik},~\bfnm{Pavel}\binits{P.}},
  \bauthor{\bsnm{Myllym{\"a}ki},~\bfnm{Mari}\binits{M.}} \AND
  \bauthor{\bsnm{Stoyan},~\bfnm{Dietrich}\binits{D.}}
(\byear{2011}).
\btitle{{Correct testing of mark independence for marked point patterns}}.
\bjournal{Ecological Modelling}
\bvolume{222}
\bpages{3888--3894}.
\end{barticle}
\endbibitem

\bibitem[\protect\citeauthoryear{Janson and Vegelius}{1981}]{Janson1981}
\begin{barticle}[author]
\bauthor{\bsnm{Janson},~\bfnm{Svante}\binits{S.}} \AND
  \bauthor{\bsnm{Vegelius},~\bfnm{Jan}\binits{J.}}
(\byear{1981}).
\btitle{Measures of ecological association}.
\bjournal{Oecologia}
\bvolume{49}
\bpages{371--376}.
\end{barticle}
\endbibitem

\bibitem[\protect\citeauthoryear{Jerrum}{2003}]{Jerrum2003}
\begin{bbook}[author]
\bauthor{\bsnm{Jerrum},~\bfnm{M}\binits{M.}}
(\byear{2003}).
\btitle{{Counting, Sampling and Integrating: Algorithms and Complexity}}.
\bseries{Lectures in Mathematics ETH Z\"{u}rich}.
\bpublisher{Birkh\"{a}user Verlag}, \baddress{Basel}.
\end{bbook}
\endbibitem

\bibitem[\protect\citeauthoryear{Jerrum and Sinclair}{1996}]{Jerrum1996}
\begin{barticle}[author]
\bauthor{\bsnm{Jerrum},~\bfnm{Mark}\binits{M.}} \AND
  \bauthor{\bsnm{Sinclair},~\bfnm{Alistair}\binits{A.}}
(\byear{1996}).
\btitle{{The Markov chain Monte Carlo method: an approach to approximate
  counting and integration}}.
\bjournal{Approximation algorithms for NP-hard problems}
\bpages{482--520}.
\end{barticle}
\endbibitem

\bibitem[\protect\citeauthoryear{Jones and Semple}{2012}]{Jones2012}
\begin{bbook}[author]
\bauthor{\bsnm{Jones},~\bfnm{Richard}\binits{R.}} \AND
  \bauthor{\bsnm{Semple},~\bfnm{Sarah}\binits{S.}}
(\byear{2012}).
\btitle{{Sense of Place in Anglo-Saxon England}}.
\bpublisher{Shaun Tyas}.
\end{bbook}
\endbibitem

\bibitem[\protect\citeauthoryear{Karpinski, Rucinski and
  Szymanska}{2012}]{Karpinski2012}
\begin{barticle}[author]
\bauthor{\bsnm{Karpinski},~\bfnm{M}\binits{M.}},
  \bauthor{\bsnm{Rucinski},~\bfnm{A}\binits{A.}} \AND
  \bauthor{\bsnm{Szymanska},~\bfnm{E}\binits{E.}}
(\byear{2012}).
\btitle{{Approximate Counting of Matchings in Sparse Uniform Hypergraphs}}.
\bjournal{arXiv preprint arXiv:1204.5335}
\bpages{1--13}.
\end{barticle}
\endbibitem

\bibitem[\protect\citeauthoryear{Kuhn}{1955}]{Kuhn1955}
\begin{barticle}[author]
\bauthor{\bsnm{Kuhn},~\bfnm{HW}\binits{H.}}
(\byear{1955}).
\btitle{{The Hungarian method for the assignment problem}}.
\bjournal{Naval Research Logistics Quarterly}
\bvolume{2}
\bpages{83--97}.
\end{barticle}
\endbibitem

\bibitem[\protect\citeauthoryear{Lau and Green}{2007}]{Lau2007}
\begin{barticle}[author]
\bauthor{\bsnm{Lau},~\bfnm{JW}\binits{J.}} \AND
  \bauthor{\bsnm{Green},~\bfnm{PJ}\binits{P.}}
(\byear{2007}).
\btitle{{Bayesian model-based clustering procedures}}.
\bjournal{Journal of Computational and Graphical Statistics}
\bvolume{16}
\bpages{526--558}.
\end{barticle}
\endbibitem

\bibitem[\protect\citeauthoryear{Lawson and Denison}{2010}]{Lawson2010}
\begin{bbook}[author]
\bauthor{\bsnm{Lawson},~\bfnm{AB}\binits{A.}} \AND
  \bauthor{\bsnm{Denison},~\bfnm{DGT}\binits{D.}}
(\byear{2010}).
\btitle{{Spatial cluster modelling}}.
\bpublisher{CRC press}.
\end{bbook}
\endbibitem

\bibitem[\protect\citeauthoryear{Loizeaux and McKeague}{2001}]{Loizeaux2001}
\begin{barticle}[author]
\bauthor{\bsnm{Loizeaux},~\bfnm{MA}\binits{M.}} \AND
  \bauthor{\bsnm{McKeague},~\bfnm{IW}\binits{I.}}
(\byear{2001}).
\btitle{{Perfect sampling for posterior landmark distributions with an
  application to the detection of disease clusters Bayesian cluster models}}.
\bjournal{Selected Proceedings of the Symposium on Inference for Stochastic
  Processes}
\bvolume{37}
\bpages{321--332}.
\end{barticle}
\endbibitem

\bibitem[\protect\citeauthoryear{Marinari and Parisi}{1992}]{Marinari1992}
\begin{barticle}[author]
\bauthor{\bsnm{Marinari},~\bfnm{E}\binits{E.}} \AND
  \bauthor{\bsnm{Parisi},~\bfnm{G}\binits{G.}}
(\byear{1992}).
\btitle{{Simulated tempering: a new Monte Carlo scheme}}.
\bjournal{EPL (Europhysics Letters)}
\bvolume{c}
\bpages{1--12}.
\end{barticle}
\endbibitem

\bibitem[\protect\citeauthoryear{M{\"u}ller and Quintana}{2010}]{Muller2010}
\begin{barticle}[author]
\bauthor{\bsnm{M{\"u}ller},~\bfnm{Peter}\binits{P.}} \AND
  \bauthor{\bsnm{Quintana},~\bfnm{Fernando}\binits{F.}}
(\byear{2010}).
\btitle{{Random partition models with regression on covariates}}.
\bjournal{Journal of Statistical Planning and Inference}
\bvolume{140}
\bpages{2801--2808}.
\end{barticle}
\endbibitem

\bibitem[\protect\citeauthoryear{Oh, Russell and Sastry}{2009}]{Oh2009}
\begin{barticle}[author]
\bauthor{\bsnm{Oh},~\bfnm{Songhwai}\binits{S.}},
  \bauthor{\bsnm{Russell},~\bfnm{Stuart}\binits{S.}} \AND
  \bauthor{\bsnm{Sastry},~\bfnm{Shankar}\binits{S.}}
(\byear{2009}).
\btitle{{Markov chain Monte Carlo data association for multi-target tracking}}.
\bjournal{Automatic Control, IEEE Transactions on}
\bvolume{54}
\bpages{481--497}.
\end{barticle}
\endbibitem

\bibitem[\protect\citeauthoryear{Plummer et~al.}{2005}]{Plummer2005}
\begin{barticle}[author]
\bauthor{\bsnm{Plummer},~\bfnm{M}\binits{M.}},
  \bauthor{\bsnm{Best},~\bfnm{N}\binits{N.}},
  \bauthor{\bsnm{Cowles},~\bfnm{K}\binits{K.}} \AND
  \bauthor{\bsnm{Vines},~\bfnm{K}\binits{K.}}
(\byear{2005}).
\btitle{{Output analysis and diagnostics for MCMC}}.
\bjournal{{R package version 0.10-3, URL http://cran. rproject. org}}.
\end{barticle}
\endbibitem

\bibitem[\protect\citeauthoryear{Roberts}{1998}]{Roberts1998}
\begin{barticle}[author]
\bauthor{\bsnm{Roberts},~\bfnm{GO}\binits{G.}}
(\byear{1998}).
\btitle{{Optimal Metropolis algorithms for product measures on the vertices of
  a hypercube}}.
\bjournal{Stochastics and Stochastic Reports}
\bvolume{62}
\bpages{275--283}.
\end{barticle}
\endbibitem

\bibitem[\protect\citeauthoryear{Roberts, Gelman and Gilks}{1997}]{Roberts1997}
\begin{barticle}[author]
\bauthor{\bsnm{Roberts},~\bfnm{Gareth~O}\binits{G.~O.}},
  \bauthor{\bsnm{Gelman},~\bfnm{Andrew}\binits{A.}} \AND
  \bauthor{\bsnm{Gilks},~\bfnm{Walter~R}\binits{W.~R.}}
(\byear{1997}).
\btitle{{Weak convergence and optimal scaling of random walk Metropolis
  algorithms}}.
\bjournal{The Annals of Applied Probability}
\bvolume{7}
\bpages{110--120}.
\end{barticle}
\endbibitem

\bibitem[\protect\citeauthoryear{Valiant}{1979}]{Valiant1979}
\begin{barticle}[author]
\bauthor{\bsnm{Valiant},~\bfnm{LG}\binits{L.}}
(\byear{1979}).
\btitle{{The complexity of enumeration and reliability problems}}.
\bjournal{SIAM Journal on Computing}
\bvolume{8}
\bpages{410--421}.
\end{barticle}
\endbibitem

\end{thebibliography}

\end{document}